\documentclass[reprint,amsmath,amssymb,aps,prl]{revtex4-2}

\usepackage{graphicx} 
\usepackage{bm} 
\usepackage{amsmath,amssymb}
\usepackage{xcolor}

\graphicspath{ {./}{figs/}}
\begin{document}

\title{First-principles Spin and Optical Properties of Vacancy Clusters in Lithium Fluoride}

\author{Mariano Guerrero Perez$^{1}$}
\author{Keegan Walkup$^{1}$}
\author{Jordan Chapman$^{2}$}
\author{Pranshu Bhaumik$^{4}$}
\author{Giti A. Khodaparast$^{1}$}
\author{Brenden A. Magill$^{1}$}
\author{Patrick Huber$^{1}$}
\author{Vsevolod Ivanov$^{1,2,3}$ }
\email{vivanov@vt.edu}
\affiliation{$^1$Department of Physics, Virginia Tech, Blacksburg, Virginia 24061, USA}
\affiliation{$^2$Virginia Tech National Security Institute, Blacksburg, Virginia 24060, USA}
\affiliation{$^3$Virginia Tech Center for Quantum Information Science and Engineering, Blacksburg, Virginia 24061, USA}
\affiliation{$^4$College of William and Mary, Williamsburg, VA 23187, USA}

\begin{abstract}
Vacancy-cluster color centers in lithium fluoride have been studied in detail both theoretically and experimentally for over a century, giving rise to various applications in solid-state lasers, broadband photonic devices, and radiation dosimeters. These color centers are also attractive candidate platforms for applications in quantum information science, due to their spin properties and strong coupling to the crystal lattice, which allows their properties to be easily tuned. Here we present hybrid functional calculations of common vacancy defects in lithium fluoride, including their energetic, spin, and optical properties. We show that for a wide range of hybrid functional parameters tuned to match the experimental band gap, certain defects have little variation in their predicted optical properties. We further demonstrate that the parameters needed to satisfy the generalized Koopman's theorem and correctly position defect levels within the gap, can vary dramatically, even for different charge states of the same defect. Our work establishes the accuracy of the computationally lightweight hybrid-functional approach for predicting the optical and energetic properties of color centers in polar materials.
\end{abstract}

\maketitle

\section{Introduction.}

After initial observations over a century ago of the discoloration of alkali halide crystals following cathode ray exposure, the subsequent decades focused on the interpretation of these results in the context of the modern theory of crystal structure and color center point defects \cite{Seitz_1946}. In addition to being of broad physical interest, lithium fluoride (LiF) received particular attention due to its attractive optical properties, such as its wide bandgap of 14.2 eV \cite{Piacentini}, optical activity throughout the visible light spectrum at room temperature \cite{Nichelatti}, and outstanding thermal stability of its color center defects \cite{Baldacchini}. Detailed studies of the optical and electronic properties of LiF color centers led to their application in a variety of technologies, both proposed and realized, including optically-pumped tunable solid-state lasers \cite{Ter-Mikirtychev}, broadband light-emitting photonic devices \cite{Montereali, Gellermann}, passive ionizing radiation dosimeters \cite{Montereali2}, and, more recently, dark matter detection \cite{PALEO}. 

The optical absorption of LiF at 250 nm after exposure to hard x-rays at room temperature was reported in 1928 by Ottmer \cite{Ottmer}. Throughout the next few decades, the optical activity of LiF at 250 nm was attributed to an electron trapped at a fluorine (F) vacancy within the crystal lattice comprising the \textit{F} absorption band \cite{Pringsheim, Klick, Uchida, Delbecq, Bate, Gilman, Kalo, Weigand, Fischer}. This picture of color centers---known as the vacancy model---in LiF gained acceptance as a result of work in 1938 by Pohl, who generally termed the characteristic absorption peaks in alkali halides as \textit{F}-centers \cite{ionic_book}, after \textit{Fabre}, the German word for color. An understanding of the photoluminescence of LiF was borne primarily from analogy to other alkali halides \cite{Kaufman}; however, observed differences in its optical behavior launched a rigorous investigation into the electronic structures of its color center defects that continued in the following decades \cite{Gorlich, Kaufman, Nahum}.

Since then, insights into color center defects in LiF have informed their potential application in quantum information sciences (QIS); color centers act as isolated molecular systems that can be exploited as single-photon emitters, magnetic field sensors, and spin-photon-entangled systems \cite{Norman, Wasielewski, Bayliss}. LiF color centers are strong candidates for QIS applications due to the inherent tunability of LiF material properties to achieve coupling of defect quantum states to measurable quantities \cite{Duarte}. The identification and screening of quantum defects with desired properties is infeasible with experimental approaches alone; thus, high-throughput \textit{ab initio} methods have been employed to accelerate the discovery and cataloging of quantum defects in user-friendly databases \cite{Ivanov-db, xiong-db, davidsson-db}.

In this work we systematically study common color center defects in LiF and provide a definitive picture of their electronic structures. In particular, we consider the applicability of the Heyd-Scuseria-Ernzerhof  \cite{HSE} hybrid functional approach commonly used in calculations of color center defects and show how tuning of HSE parameters allows us to reproduce the material properties of LiF in a fully self-consistent manner while accurately estimating relevant defect formation energies. A hybrid functional approach was chosen for its computational affordability, enabling supercell calculations large enough to minimize errors from interacting periodic defect images. 


This paper is organized as follows: we begin by discussing the historical results, from both theory and experiment, that established the electronic structures of each of the LiF color centers and compare them with our calculations. We focus on calculations of both the $F$-center and aggregate $F$-centers that have been assigned previously to the absorption and emission peaks in LiF. We present new insights into the significance of the symmetry of the color center defects, including symmetry breaking structural distortions that break the degeneracy of the $F$-center and $F_3^+$-center excited states, as well as a spin-dependent Jahn-Teller effect in aggregate $F$ defects. Furthermore, we discuss the tuning protocol for HSE parameters that satisfy the generalized Koopmans' theorem (gKT) and provide critical consideration of the utility of these solutions for multi-atom defects in LiF. Finally, we provide our computational description of the electronic structures of the excited defect states that are ascribed to the optical activity of LiF with particular attention paid to the strong electron-phonon coupling in LiF. 

\section{Color centers in LiF.}

The observation that LiF, unlike other alkali fluorides, could not be activated optically through additive coloring or light bleaching necessitated more in-depth studies of the nature of its color centers \cite{Gorlich}. Despite the reproducible absorption band near 250 nm in LiF crystals, experimental evidence that this optical activity was due to an $F$ center in LiF was still scant by the early 1960s. For instance, Kaufman \textit{et al.} demonstrated that optical bleaching of the 250-nm absorption band in LiF with "\textit{F}" light (253.7 nm) led to the eventual loss of its optical activity \cite{Kaufman}. Optical bleaching of other alkali halides with the same cubic lattice structure resulted in the formation of aggregate color centers \cite{Duerig, Petroff}, suggesting that the analogous low-wavelength absorption band in LiF could be distinct from \textit{F} bands in other alkali halides. 

Nonetheless, efforts to characterize the \textit{F} center and aggregate \textit{F} centers---color center defects composed of more than one F vacancy, in the case of LiF---were prolific in the following decades. LiF was previously shown to be optically active at higher wavelengths, having absorption bands at 320 nm, 378 nm, and 450 nm \cite{Ivey}. Although the location of the \textit{F}-band in LiF could not be definitively established, the 450-nm absorption peak was confirmed as the $M$ band \cite{Kaufman}. Earlier work provided evidence that optical activity the $M$ band originated from an $M$-center formed from two associated $F$-centers \cite{vanDoorn, Faraday}, i.e., an $F_2$ center. Excitation in the $M$ region was found to produce green and red emission bands at 528 nm and 670 nm, respectively \cite{Okuda}. The $M$ absorption band, however, was also found to obscure the optical signatures of four weaker bands in the $M$ range but not associated with the $F_2$ center of LiF \cite{Nahum}. The green emission band at 528 nm is instead theorized to originate from optical transitions at an ionized $R$ center, thought to be two electrons trapped at an $F_3$-center \cite{Nahum}.

\begin{figure}[t]
    \includegraphics[width=1.0\columnwidth]{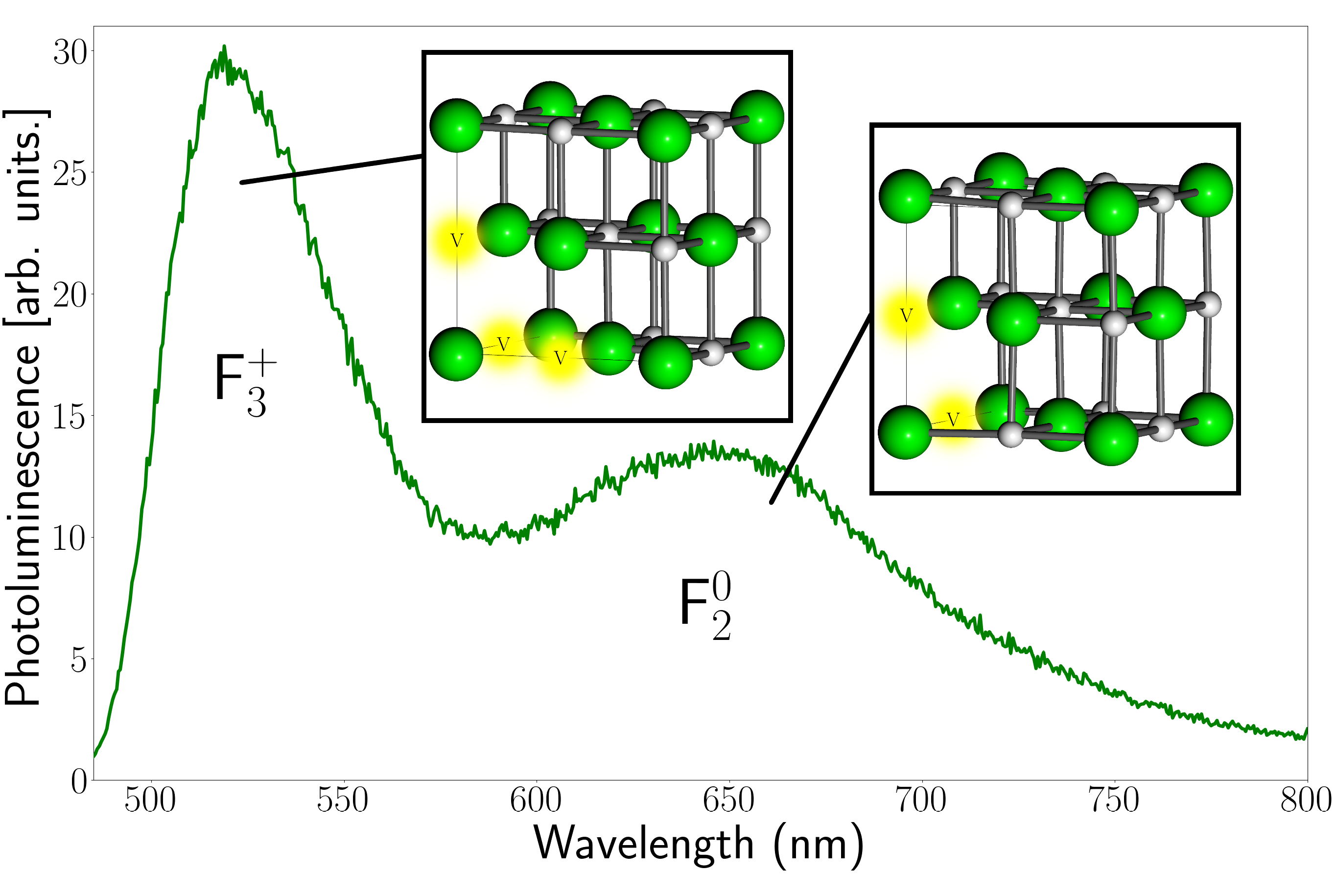}
    \caption{Photoluminescence spectrum of LiF, showing two broad emission peaks at $\sim$525 nm and $\sim$650 nm corresponding to the $F_2^0$-center and $F_3^+$-center respectively, with insets showing their structures.}
    \label{pl}
\end{figure}

Figure \ref{pl} shows the photoluminescence spectrum of LiF, displaying the characteristic wide $M$ and $R$ emission peaks, now known to be associated with the neutral $F_2$ and positively charged $F_3^+$ defects \cite{Baldacchini_LiF_seasons}. For the spectra, 10 mm$\times$10 mm$\times$10 mm LiF cubes with six optically polished sides were obtained from Crystran. These were irradiated at room temperature for 45 minutes with gamma rays from a $^{60}$Co source with an approximate fluence of 3$\times 10^{10}$ photons per cm$^2$ to generate $F_2$ and $F_3^+$ vacancy clusters. Photoluminescence spectra were taken prior and post irradiation using an Agilent Cary Eclipse spectrometer with an excitation wavelength of 450 nm.

We now proceed with a detailed discussion of each type of color center, and a comparison of the computed electronic structures with established defect characteristics.

\begin{figure*}[t]
    \includegraphics[width=1.0\textwidth]{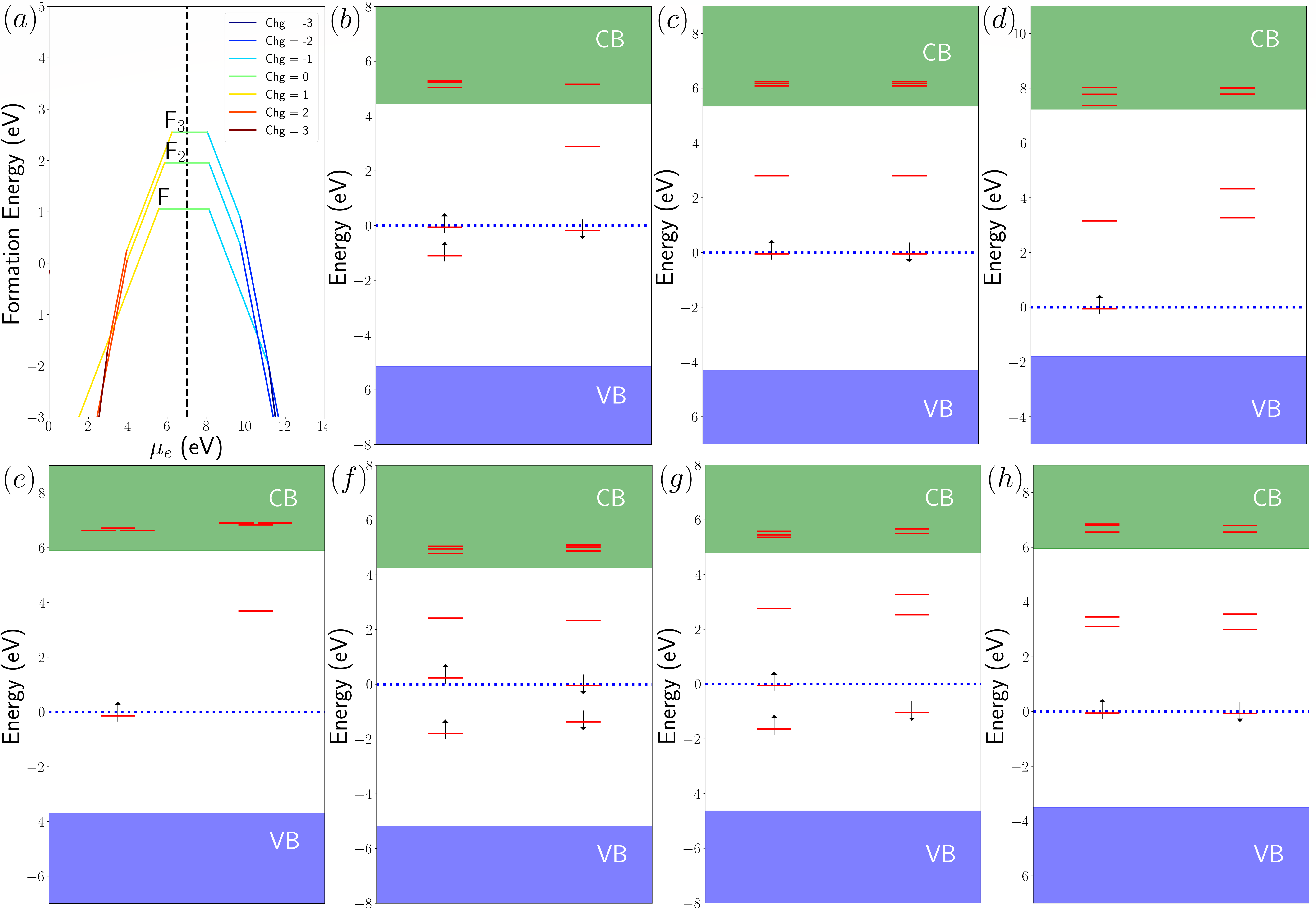}
    \caption{Electronic structures of vacancy centers in LiF. Formation energies as a function of chemical potential (a) for each defect/charge state. Defect levels (red) relative to the conduction band (CB, green), and valence band (VB, blue) of LiF for $F_2^-$ (b), $F_2^0$ (c), $F_2^+$ (d), $F^0$ (e), $F_3^-$ (f), $F_3^0$ (g), and $F_3^+$ (h). }
    \label{lvls}
\end{figure*}

\subsection{$F$-center}



Despite extensive experimental efforts \cite{Seitz_1954, Baldacchini_LiF_seasons}, the luminescence signal of the monovacancy $F$-center has not been observed. Nevertheless, the electronic structure of the $F$-center has been studied extensively at various levels of theory \cite{Chaney_LCAO, Pederson, Mallia, Komel_F_M_embedded_cluster, Ewig_cluster, Blaha_Fcenter}. Early models based on linear combinations of atomic orbitals \cite{Chaney_LCAO}, captured the essential features of the electronic structure - a non-degenerate ground state, and a triply-degenerate excited state, with a computed absorption of 4.85 eV. Density functional theory (DFT) calculations based on the local density approximation (LDA) corroborated this model, placing the occupied $s$-like a$_\text{1g}$ defect state approximately 9.0 eV above the valence band maximum (VBM). An unoccupied $p$-like t$_\text{1u}$ defect state was found to be 4.90 eV higher in energy, lying just above the conduction band minimum (CBM) \cite{Pederson}, in reasonable agreement with experimental absorption values of $\sim$ 5.1 eV. As defect levels tend to hybridize with bulk states, the computed positions of the VBM and CBM can have a significant effect on the defect electronic structure, particularly for LDA, which is known to underestimate insulator band gaps \cite{DFT-gaps,HSE}. More recent work on the $F$-center has incorporated electronic interactions to remedy this issue. Incorporating Hartree-Fock (HF) exchange has been show to be critical for describing the expected localization of in-gap defect states near the band edges and increases the separation between the CBM and the a$_\text{1g}$ defect state to 11 eV \cite{Mallia}. 

G$_0$W$_0$ \cite{Chen} and GW+BSE \cite{Blaha_Fcenter} calculations have also been performed for the $F$-center, improving the bandgap prediction to within 1 eV of the experimental value. When comparing G$_0$W$_0$ with HSE, the defect level as predicted by HSE calculations (8.59 eV above VBM) was found to sit about 0.12 eV above that calculated for G$_0$W$_0$ theory (7.84 eV above VBM); however, the defect level predicted by the G$_0$W$_0$ scheme was found to lie nearly 0.8 eV deeper within the band gap. The computational complexity of these methods is a significant drawback, limiting their application to very small supercell sizes ($\sim$ 30-60 atoms), and requiring various finite-size corrections and extrapolations to account for defect image interactions \cite{Chen, Blaha_Fcenter}. Quantum chemistry approaches using a finite real-space unit cell that avoids issues related to periodic images, have also been successful at reproducing the electronic structure of the $F$-center \cite{Komel_F_M_embedded_cluster, Ewig_cluster, Blaha_Fcenter}. Within the quantum chemical method, the ground-state and excited-state wave functions of the $F$-center defect are calculated after explicit placement of the trapped electron in the $s$- or $p$-state, corresponding to the ground state and excited state, respectively. Using this method yields a band gap of 14.5 eV, closely matching the experimental gap for LiF, and reproduces the both $s$-like ground state and $p$-like excited state of the defect within the band gap \cite{Blaha_Fcenter}.

To investigate the vacancy-related color center defects in LiF, we adapt the high-throughput workflow that has been used successfully to reproduce the optical properties of color centers in various other materials \cite{Ivanov-db}. All calculations were performed using the Vienna \textit{ab initio} Simulation Package (VASP) with the projector augmented wave (PAW) method \cite{Kresse_potentials}. Color center defects are embedded into a $3 \times 3 \times 3$ LiF supercell containing 216 atoms, with the experimental lattice parameter of 4.0263 \AA ~\cite{Piacentini}, and a $\Gamma$ k-point mesh. All considered structures are relaxed to a force tolerance of $10^{-3}$ \AA/eV.  Visualization and structure editing were done in the VESTA software package \cite{Momma}, and post-processing calculations were performed using VASPKIT \cite{VASPKIT}.

It is well known that atomic displacements in the presence of a defect can have a significant effect (up to several eV) on the electronic structure \cite{Chaney_LCAO, Pederson, Blaha_Fcenter}. In our calculation, we allow for the full relaxation of the atomic structure, without constraining to the cubic symmetry commonly assumed for the $F$-center. We find that the structure undergoes a small energy-reducing axial distortion, resulting in the splitting of the t$_\text{1u}$ defect level into an a$_\text{2u}$ level and a doubly degenerate e$_\text{u}$ level as shown in Fig. \ref{lvls}(e). In the occupied spin channel, the e$_\text{u}$ states lie 81 meV higher in energy, while in the other spin channel, the a$_\text{2u}$ state is much more delocalized and lies 59 meV above the doubly degenerate state. The strong electron-phonon coupling would likely lead to broadening in the emission peak that obscures this symmetry breaking; however, at very low temperatures, the emission peaks may be narrow enough to resolve the splitting. 


\subsection{$F_2$-center}

Early work on the $F_2$-center ($M$-center) established that its structure consists of two adjacent $F$-centers on nearest neighbor sites \cite{Meyer_F2}. This was further corroborated by measurements of the relative abundance of $F_2$ and $F$ centers, which showed that the concentration of $F_2$ varied quadratically with the concentration of $F$ centers \cite{Faraday_F2}. The $F_2$ centers readily captured and emitted electrons, forming the charged $F_2^+$ and $F_2^-$. Switching between these charge states could be readily achieved by either light \cite{Tsuboi_F2_oscillation} or electron pulses \cite{Lisitsyna_EP}, and their emission bands were observed to have high quantum efficiencies, even at room temperature enabling high-power applications \cite{Baldacchini_LiF_seasons}. 

Our first principles calculations of the $F_2$-center and its charged counterparts, reveal two defect levels per spin-channel within the band gap (Figs. \ref{lvls}(b),(c),(d), along with several more non-degenerate unoccupied states above the CBM. This is consistent with electronic models of the $F_2$ center which predict a number of non-degenerate ground and excited states \cite{Meyer_F2}.

Time resolved measurements of the neutral $F_2$-center have also revealed a long-lived metastable triplet state that is thought to form through the localization of a conduction electron onto an $F_2^+$-center \cite{Lisitsyna1994_F2_triplet}. To compare, we also computed the $F_2$ triplet ground state, placing it $290$ meV above the ground state singlet. The triplet state has a moderate zero-field splitting (ZFS) of $\sim 1.6$ GHz.

\subsection{$F_3$-center}

The $F_3$-center ($R$-center) in LiF has two closely placed absorption bands around $\sim 320$ nm and $\sim 380$ nm, respectively denoted $R_1$ and $R_2$. The constant intensity ratio of these bands, irrespective of synthesis conditions, led to their interpretation as arising from the same color center defect. Polarization measurements and the double degeneracy of the excited state of ionized state led to the identification of a three-fold symmetry in this defect \cite{Okuda_R111} - a set of three nearest-neighbor vacancies in the $\langle 111 \rangle$ plane (Fig. \ref{pl}). A related defect with an absorption band that completely overlapped with that of the $F_2$-center \cite{Baldacchini_LiF_seasons}, was also quickly identified as the ionized $F_3^+$ counterpart to this defect. The negatively charged $F_3^-$ defect was also observed, coexisting with $F_2^-$, whose absorption band overlapped with the $F_3^-$ emission \cite{Duarte_laser}, later becoming the focus of efforts to produce high-power color center lasers \cite{Tsuboi1999_F3m}.

Calculations of the defect states suggested a singlet ground state and excited state $e$ doublet for both $F_3^0$ \cite{Wang_F3_states} and $F_3^+$ \cite{Sati_F3p} defects. Our computed electronic structures for the three $F_3$ charge states (Figs. \ref{lvls}(f),(g),(h) lack this double degeneracy, due to a breaking of the three-fold symmetry of the defect. 

Later, careful temperature-dependent optical measurements identified a metastable triplet state connected to the excited singlet state through non-radiative relaxation processes \cite{Baldacchini_F3p_triplet}. For comparison, we computed the triplet states of the $F_3^+$ as well as the $F_3^-$ defects (See Supplementary Material \cite{supplement} for details). The triplet state of the $F_3^+$ is about 0.9 eV higher in energy than the singlet ground state, while for the $F_3^-$ we find that the triplet ground state is $266$ meV lower in energy, and should be energetically preferable. In a later section, calculated emission frequencies and lifetimes of these triplet states will be presented, which suggest that they may be responsible for the observed emission peaks of $F_3^-$ and $F_3^+$. 


\section{HSE tuning of the LiF band gap.}

As discussed previously, hybridization between the bulk bands and localized defect levels leading to defect states that are effectively pinned near the VBM or CBM, meaning that accurate calculations of electronic transitions are underlaid by the accurate prediction of the host band gap. This issue is particularly prominent for DFT, which is known to systematically underestimate insulator band gaps \cite{DFT-gaps, DFT-gaps2}, but can be remedied in hybrid functional approaches by a judicious choice of parameters \cite{Deak, Deak2, Ivady_screening, Blaha_Fcenter}.


The procedure to reproduce the experimental band gap of LiF was as follows. The initial geometry relaxation of pristine LiF was calculated at the Perdew-Burke-Ernzerhof (PBE) level of theory within the generalized gradient approximation (GGA) \cite{Kresse_GGA} at a plane-wave cutoff energy of 400 eV. Relaxation of the crystal lattice proceeded until the forces converged to 1 meV/\AA. We implemented HSE-level calculations to further relax the ionic geometry and elucidate the optimized electronic structure of LiF. Relaxation at the HSE level of theory was calculated with a plane-wave cutoff energy of 520 eV to a tolerance of 1 meV/\AA. 

\begin{figure}[htb]
    \includegraphics[width=1.0\columnwidth]{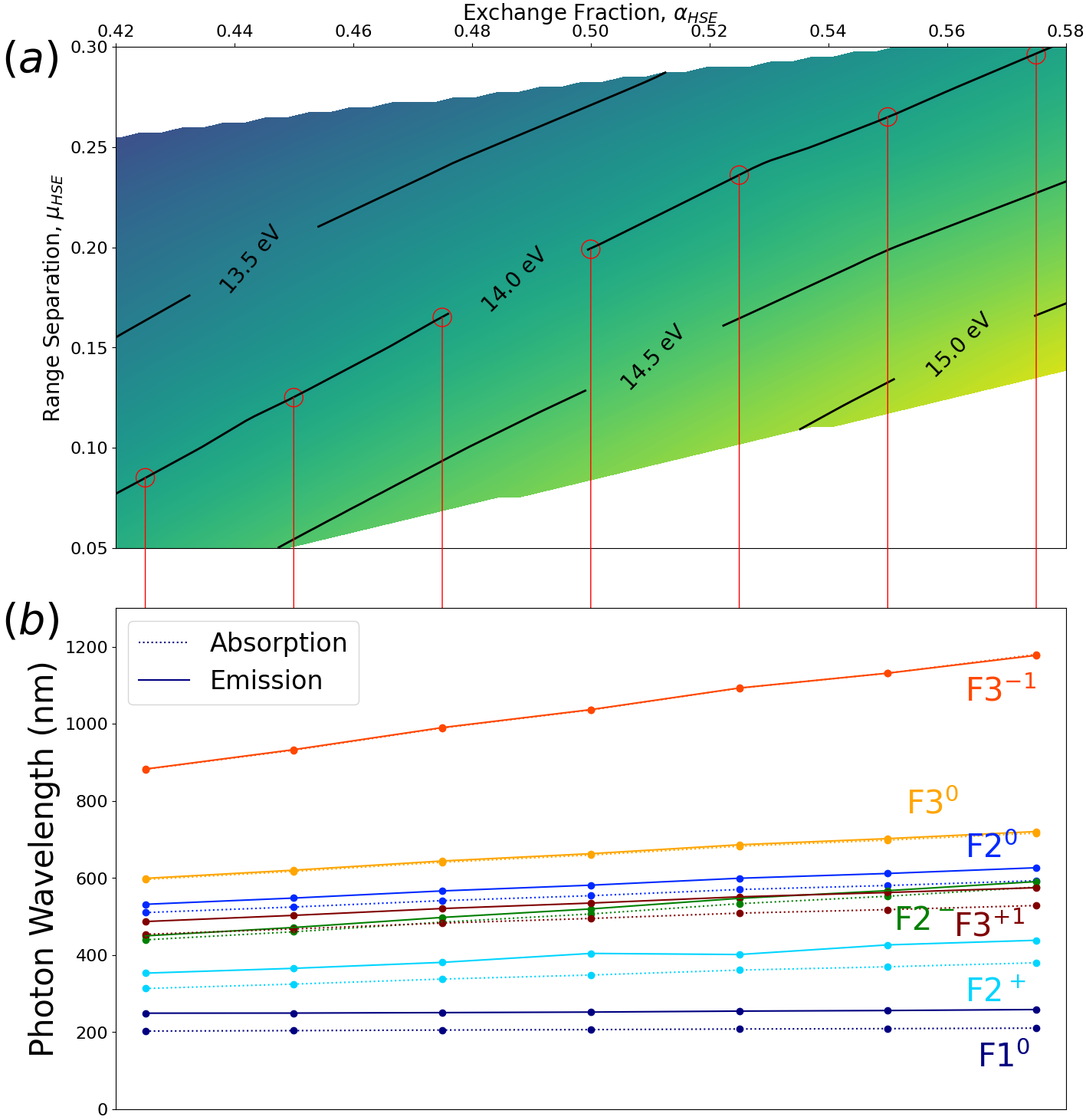}
    \caption{Contour plot (a) showing the band gap of LiF as a function of $\alpha_\text{HSE}$ and $\mu_\text{HSE}$. Red circles indicate the points along the 14.0 eV isoline for which defect optical properties were computed. (b) Absorption (dashed) and emission (solid) for each color center defect as a function of HSE parameters along the 14.0 eV isoline in (a). }
    \label{iso}
\end{figure}

The HSE formalism is defined by two parameters, the proportion of HF exact exchange to be contributed to the hybrid functional, $\alpha_\text{HSE}$, and the range separation parameter, $\mu_\text{HSE}$. A popular choice is the HSE06 functional ($\alpha_\text{HSE}=0.25$, $\mu_\text{HSE}=0.2$), which accurately reproduces experimental band gap in a wide range of materials \cite{HSE, Deak}. Furthermore, the $1/4$ exact exchange fraction has been shown to correctly mimic correlation interactions for materials that are well-described by a unpolarized electron gas \cite{Bernardi_one_fourth}, but there is no reason to expect that this choice of HSE parameters is suitable for polar materials. In fact, our calculations of LiF with the HSE06 functional result in a band gap of 11.6 eV, significantly underestimating experimental measurements. A pure HF treatment, on the other hand, has been shown to severely overestimate the band gap of LiF \cite{Blaha_Fcenter}.

The band gap of LiF was systematically computed for $0.4 < \alpha_\text{HSE} < 0.6$ and $0.05 < \mu_\text{HSE} < 0.3$ (Figure \ref{iso}a). Iso-energetic contour lines show the $\alpha_\text{HSE}$/$\mu_\text{HSE}$ combinations that yield particular band gap energies. These isolines are nearly linear and gently sloped, indicating that the increased contribution from HF exact exchange requires only slight tuning of the complementary range separation parameter to maintain the band gap. Points were selected along the isoline that reproduced the experimental band gap of LiF (approximately 14.0 eV), for which defect optical properties would be computed. Further tuning of the HSE parameters in later parts of this work is done along this same 14.0 eV isoline, in order to maintain agreement with the experimental band gap.




\begin{figure}[htb]
    \includegraphics[width=1.0\columnwidth]{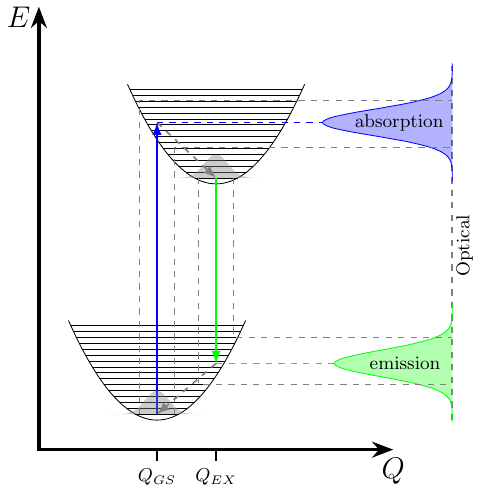}
    \caption{Diagram of the ground state ($q_{GS}$) and excited state ($q_{EX}$) configuration phonon levels, with arrows indicating optical absorption (blue), optical emission (green), and non-radiative relaxation (dashed grey). Dashed lines show the range of possible excitations that lead to broadening of emission/absorption peaks (left).}
    \label{el-ph}
\end{figure}

\subsection{Details of optical properties}

Excited electronic states are handled using the constrained occupation ($\Delta$-SCF) method, that has been well established for computing color center optical transitions \cite{Gali-delta-SCF}. While for many color centers it is sufficient to compute the energy difference between fully relaxed ground and excited states \cite{Ivanov_2, Ivanov-db}, electronic transitions in LiF require a more careful treatment on account of the strong electron-phonon coupling, that manifests as broad, Stokes-shifted peaks in the photoluminescence spectrum \cite{Montereali3}. Figure \ref{el-ph} shows a schematic Franck-Condon diagram of the electronic excitation and emission process of a defect in LiF. Upon absorbing a photon in its relaxed ground state atomic configuration, $q_\text{GS}$, the electron undergoes a vertical excitation process to a higher vibrational level of the excited state, which is followed by a non-radiative relaxation to the lowest energy excited state atomic configuration, $q_\text{EX}$. Subsequently, the system undergoes a vertical emission, followed by a final non-radiative relaxation of the geometry to the original $q_\text{GS}$ configuration. The vertical absorption/emission processes occur much faster than the ionic relaxation via nuclear vibrations. To approximate these two processes, two separate calculations of the electronic ground and excited states are performed, using the relaxed ground ($q_\text{GS}$) and excited ($q_\text{EX}$) geometries respectively \cite{supplement}. 



\begin{figure}[hb]
    \includegraphics[width=1.0\columnwidth]{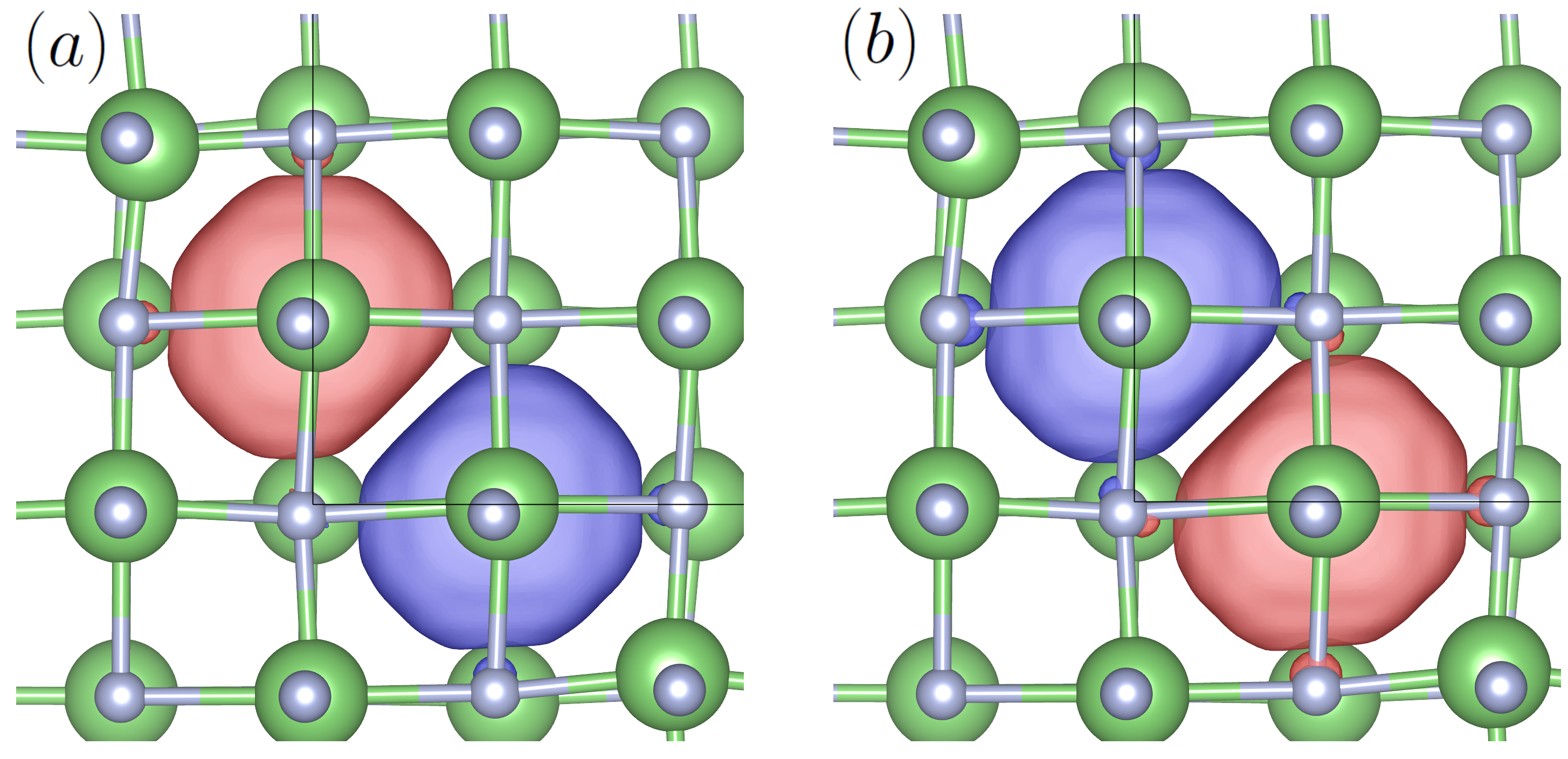}
    \caption{Plots of $|\Psi|^2$ for the (a) higest occupied and (b) lowest unoccupied defect levels of the $F_2$-center, in the spin up (red) and spin down (blue) channels. }
    \label{jt}
\end{figure}

\subsection{Spin-dependent Jahn-Teller instability}

Strong couplings between degenerate orbitals in color center defects are known to lead to a spontaneous symmetry breaking, known as the dynamical Jahn-Teller effect \cite{Gali-JT1, Gali-JT2}. Coupling of the defect states to local phonon-modes results in a splitting of the degenerate states, and an overall reduction in energy. This effect poses a particular challenge for excited state calculations using the constrained occupation method \cite{Gali-delta-SCF}, since integer excitations into one orbital of a degenerate set leads to an instability that prevents electronic convergence. This issue can be avoided by a crude representation of the correlated state using half-electron excitations into each orbital \cite{Gali-JT1}.

When performing $\Delta$-SCF calculations of the excited states for the $F_2$-center and $F_3$-center, a similar issue arises. Figure \ref{jt} shows the occupied and unoccupied defect levels within the gap for the $F_2$-center. A spin-conserving excitation in a single spin channel results in a charge transfer from one vacancy site to another, leading to a charge transfer and an instability that prevents electronic convergence. Just as for degenerate states, the instability is avoided by performing the excited state calculation with a half-electron excitation in each spin channel, mimicking a spin-conserving singlet excitation \cite{Gali-JT1}.

\subsection{Optical properties along the isoline}



The absorption and emission of each defect center in LiF are plotted in Figure \ref{iso}b for each point on the 14 eV band gap isoline. Despite their physical similarity, the different vacancy cluster defects have varying degrees of dependence on the HSE parameters. As the isoline is necessarily chosen to consistently reproduce the properties of bulk LiF, a stronger dependence on the particular HSE parameters then implies that the behavior of the defect is more decoupled from the host lattice. Indeed, experimental observations indicate that $F$ and $F_3^0$ defects have very strong coupling to the LiF lattice \cite{Baldacchini_LiF_seasons}, which is consistent with their nearly constant optical properties along the isoline. The $F_3^+$ and $F_3^-$ centers, conversely, are believed to have very weak coupling to the lattice \cite{Baldacchini_LiF_seasons}, consistent with their sensitivity to the HSE parameters.

\begin{table*}[ht]
    \begin{tabular}{ c c c c c | c c c c c c c c c c }
        \hline
        \hline
        Defect & $\alpha_\text{HSE}$ & $\mu_\text{HSE}$ & $\Delta_\text{HOMO}^{(N)}$ & $\Delta_\text{LUMO}^{(N-1)}$ & $\bm{S}$ & $E_\text{a}$(eV) & $E_\text{e}$(eV) & $\lambda_\text{a}$(nm) & $\lambda_\text{e}$(nm) & $\lambda^\text{EX}_\text{a}$(nm) & $\lambda^\text{EX}_\text{e}$(nm) & TDM$^2$ (Debye$^2$) & $\tau$(ns) & $\tau^\text{EX}$(ns)\\
        \hline
        F$^0$   & 0.450 & 0.125 & -5.97 & -5.96 & $0$   & 6.078 & 4.969 & 204 & 250 & 250 & N/A & 1.7 & 42  & \\
        F$_2^-$ & 0.697 & 0.473 & -5.31 & -5.31 & $1/2$ & 1.359 & 1.183 & 912 & 1048& 956 & 1113& 14.5& 364 & \\
        F$_2^0$ & 0.381 & 0.029 & -6.05 & -6.05 & $0$   & 2.475 & 1.909 & 501 & 649 & 443 & 678 & 2.6 & 490 & 17 \\
                &       &       &       &       & $1$   & 4.247 & 4.013 & 292 & 309 & & & 1.2 & 114 & \\
        F$_2^+$ & 0.383 & 0.031 & -7.47 & -7.46 & $1/2$ & 4.063 & 3.895 & 305 & 318 & 645 & 910 & 6.5 & 23 & 29 \\
        F$_3^-$ & 0.556 & 0.275 & -3.73 & -3.73 & $0$   & 1.955 & 1.949 & 634 & 636 & 795 & 898 & 1.7 & 703 & 10 \\
                &       &       &       &       & $0$   & 1.953 & 1.900 & 635 & 653 & & & 5.6 & 226 & \\
                &       &       &       &       & $1$   & 1.979 & 1.646 & 627 & 753 & & & 7.9 & 245 & \\
        F$_3^0$ & 0.403 & 0.060 & -5.10 & -5.10 & $1/2$ & 2.157 & 2.144 & 575 & 578 & & & 0.7 & 1337 & 50 \\
                &       &       &       &       & $1/2$ & 3.715 & 3.470 & 334 & 357 & 380 & & 3.5 & 56 & \\
                &       &       &       &       & $1/2$ & 3.862 & 3.394 & 321 & 365 & 310 & & 3.1 & 71 & \\
        F$_3^+$ & 0.368 & 0.010 & -7.15 & -7.14 & $0$   & 2.815 & 2.631 & 440 & 471 & 443 & 542 & 5.4 & 90 & 11.5\\
                &       &       &       &       & $1$   & 2.493 & 2.204 & 497 & 563 & & & 0.5 & 1607 & \\
                &       &       &       &       & $1$   & 4.385 & 3.900 & 283 & 318 & & & 5.2 & 28 & \\
        \hline
        \hline
    \end{tabular}
    \caption{HSE parameters $\alpha_\text{HSE}$ (exact exchange fraction) and $\mu_\text{HSE}$ (range separation parameter) satisfing generalized Koopman's theorem for each defect, and the resulting defect level positions ($\Delta_\text{HOMO}$, $\Delta_\text{LUMO}$) relative to the conduction band. The spin states $\bm{S}$, absorption/emission energies ($E_\text{a}$, $E_\text{e}$)/wavelengths ($\lambda_\text{a}$, $\lambda_\text{e}$) compared with experimental values \cite{Baldacchini_LiF_seasons} ($\lambda^\text{EX}_\text{a}$, $\lambda^\text{EX}_\text{e}$), transition dipole moments (TDM), and emission radiative lifetimes ($\tau$, $\tau^\text{EX}$ \cite{Bosi_lifetimes}) are given for each electronic transition. }
    \label{defectdata}
\end{table*}

Even when the optical transition remains quite consistent for each ($\alpha_\text{HSE}$, $\mu_\text{HSE}$) pair along the isoline, for certain defects, the position of localized states within the gap can still vary. For instance, we find that the position $s$-like defect state of the $F$-center \cite{Chen, Blaha_Fcenter} varies from 7.30 eV to 7.98 eV above the VBM. This range of defect locations is in good agreement with those predicted by G\textsubscript{0}W\textsubscript{0} calculations; tuning of the HSE parameters seemingly addresses the overestimation of the defect level energy within the LiF band gap predicted by the HSE06 scheme \cite{Chen, HSE}. However, the unoccupied $p$-like defect states of the $F$-center are still located above the CBM, which is a known shortcoming of DFT and hybrid DFT calculations \cite{Blaha_Fcenter}. 

\section{Generalized Koopman's Theorem}

Although the previous section established that the optical properties of many defects in LiF have little dependence on the HSE parameters (given that the band gap is correctly reproduced), a truly predictive methodology necessitates some self-consistent constraint on these parameters. One such possible constraint is known as the generalized Koopmans' theorem (gKT) \cite{Deak}, which states that the energy of the highest occupied Kohn-Sham orbital, the energy of the lowest unoccupied Kohn-Sham orbital in the ionized state, and the vertical ionization energy should all be equal \cite{Deak2}. This results in an exact cancellation of errors due to the self-interaction energy and the energy contribution from wave function relaxation, leading to the correct linear behavior \cite{Lany} of the energy as a function of fractional occupation \cite{Zunger_Koopmans}. 

Practically, the vertical ionization energy along the 14 eV isoline, chosen to reproduce the LiF bandgap, expectedly remains constant, making it impossible to achieve perfect equality of all three energies to satisfy gKT. Instead, the HSE parameters are tuned along the isoline to align the position of the highest occupied molecular (defect) orbital relative to the CBM in the $N$ electron defect system ($\Delta_\text{HOMO}^{(N)}$) and the position of the lowest unoccupied molecular (defect) orbital relative to the CBM in the ionized $N-1$ electron state ($\Delta_\text{LUMO}^{(N-1)}$) \cite{Deak2}. Proper alignment of the defect orbitals also requires a charge correction to the defect level energy, $\epsilon^\text{level}_\text{corr} = -2E^\text{tot}_\text{corr}/q$ \cite{Chen}, expressed in terms of $E^\text{tot}_\text{corr}$, the charge correction to the defect formation energy. In this way, the HSE parameters that properly align the defect levels also inherently result in the optimal formation energies.



The optimized HSE parameters satisfying gKT for each defect are given in Table \ref{defectdata}, along with the resulting $\Delta_\text{HOMO}^{(N)}$/$\Delta_\text{LUMO}^{(N-1)}$ defect level positions. For each defect, the spin-conserving absorption/emission transitions are given, along with the computed transition dipole moments and lifetimes. Known experimental values \cite{Baldacchini_LiF_seasons, Bosi_lifetimes, Kurobori_lifetimes} are shown for comparison. Further details are provided in the Supplemental Material \cite{supplement}.

To our knowledge, no Koopmans'-tuned hybrid functional calculations of the divacancy and trivacancy in LiF exist. However, our computed values of $\alpha_\text{HSE} = 0.45$ and $\mu_\text{HSE} = 0.125$ that satisfy gKT for the $F$-center, compare well with prior results, which found a solution at $\alpha_\text{HSE} = 0.47$ \cite{Chen}. We also note that the positive and neutral defect solutions have rather small values of $\mu_\text{HSE} < 0.1$, which are necessary to compensate the reduced $\alpha_\text{HSE}$ in order to satisfy the 14.0 eV bandgap constraint.




With the exception of the $F_2^+$ and $F_3^-$ defects, our computed absorption/emission wavelengths compare well with experimental values to within $\sim 100$ nm. Computed lifetimes are significantly overestimated, likely because we only consider the emission process, and no other decay mechanisms of the excited state. The experimental absorption and emission peaks for $F_3^-$ differ by $\sim 100$ nm, which lines up well with the absorption/emission difference of the triplet. The triplet transition is also optically brightest for $F_3^-$, with a TDM$^2$ of 7.9 Debye$^2$, and the ground state triplet is $\sim 250$ meV lower in energy than the ground state singlet \cite{supplement}, suggesting this state may be responsible for the experimentally observed emission peak. The results also provide a clarification of the role of the triplet state in $F_3^+$. Experiments suggest that the relaxation from the excited state can occur through a non-radiative intersystem crossing to the ground state triplet, but also posit the involvement of a triplet excited state with a lifetime of a few microseconds \cite{Baldacchini}. Our computed lifetime of 1.6 $\mu$s and emission wavelength of $\sim 560$ nm imply that the excited singlet decays non-radiatively to the \textit{excited} triplet state that then undergoes a vertical emission process, producing an emission peak around $\sim 540$ nm. 

The absorption/emission data also hold some information about the electron-phonon coupling of the defects. As can be seen in Figure \ref{el-ph}, the difference between absorption and emission energies for a defect depends both on the strength of electron-phonon coupling and the degree of localization of the defect state. Since the computed localization for each defect is quite consistent, the separation between absorption and emission energies is effectively an indicator of the electron phonon coupling. This suggests that the $F_2^-$, $F_2^0$ singlet, as well as the  $F_3^-$/$F_3^+$ triplet transitions would be expected to have enhanced Huang-Rhys factors. 

\section{Discussion.}

We have presented a detailed analysis of the electronic structure, spin, and optical properties of vacancy color centers in LiF at the level of hybrid functional theory, which is sufficiently lightweight to allow for extensive structural relaxation and exploration of possibly symmetry breaking distortions. It is demonstrated that a previously unreported symmetry-breaking structural distortion leads to a small splitting of the $p$-like excited state in the $F$-center, and a splitting of the doubly degenerate $^1e_1$ levels of $F_3^+$. We have also shown that through optimized choices of HSE parameters that satisfy gKT and match the experimental band gap of LiF, as well as careful treatment of the excited state occupations to avoid spin-dependent Jahn-Teller-like distortions, the optical transitions of most LiF color centers can be reproduced within an error of $\sim 100$ nm. 

Furthermore, we have shown that for neutral and positively charged defects in LiF, the optical properties have a weak dependence on the choice of HSE parameters, so long as the experimental band gap is reproduced. Future work will explore whether this trend holds for other alkali halides or polar materials in general, which would enable rapid screening of optically active defects in these types of materials.

Finally, we have presented evidence that the $S=1$ state of the $F_3^-$ and $F_3^+$ defects are likely responsible for the observed emission from these defect centers. Additionally, computed tensors for the triplet ground states show large zero-field splittings. These spin-$1$ color centers along with several other spin-$1/2$ defects in LiF, establish this material as a candidate spin-photon interface. 


For applications in QIS and particle detection, it is also essential to understand the detailed energetic properties of defects, particularly formation energies. In this case, it becomes necessary to select hybrid functional parameters that fulfill the generalized Koopman's theorem, leading to a proper cancellation of the self-interaction error due to insufficient screening of Coulomb interactions between localized electrons. Satisfying gKT properly positions localized defect levels relative to the band edge, which tends to improve agreement with experiment for predicted charge and optical transitions \cite{Ivady_screening}. It also leads to an improvement in the total energy calculations needed for computing formation energies that are crucial for predicting damage tracks from nuclear recoils for particle detection \cite{PALEO}.

The vacancy-cluster color center defects in LiF are easy to synthesize, and range from very weak to strong coupling to the host lattice, making them easily tunable. They are optically bright and have a variety of stable and meta stable spin-$1/2$ and spin-$1$ states. All of these properties make LiF color centers an attractive new platform to explore for quantum applications. 


\section{Acknowledgments.}
\begin{acknowledgments}
This work has been supported through the U.S. National Science Foundation Growing Convergence Research Grant OIA-2428507, titled ``Collaborative Research: GCR: Mineral Detection of Dark Matter'' and by the National Nuclear Security Administration Office of Defense Nuclear Nonproliferation R\&D through the ``Consortium for Monitoring, Technology and Verification'' under award number DE-NA0003920.
The authors acknowledge Advanced Research Computing at Virginia Tech (arc.vt.edu) for providing computational resources and technical support.
\end{acknowledgments}

\bibliographystyle{apsrev}
\bibliography{references}

\begin{thebibliography}{77}
\expandafter\ifx\csname natexlab\endcsname\relax\def\natexlab#1{#1}\fi
\expandafter\ifx\csname bibnamefont\endcsname\relax
  \def\bibnamefont#1{#1}\fi
\expandafter\ifx\csname bibfnamefont\endcsname\relax
  \def\bibfnamefont#1{#1}\fi
\expandafter\ifx\csname citenamefont\endcsname\relax
  \def\citenamefont#1{#1}\fi
\expandafter\ifx\csname url\endcsname\relax
  \def\url#1{\texttt{#1}}\fi
\expandafter\ifx\csname urlprefix\endcsname\relax\def\urlprefix{URL }\fi
\providecommand{\bibinfo}[2]{#2}
\providecommand{\eprint}[2][]{\url{#2}}

\bibitem[{\citenamefont{Seitz}(1946)}]{Seitz_1946}
\bibinfo{author}{\bibfnamefont{F.}~\bibnamefont{Seitz}}, \bibinfo{journal}{Rev.
  Mod. Phys.} \textbf{\bibinfo{volume}{18}}, \bibinfo{pages}{384}
  (\bibinfo{year}{1946}),
  \urlprefix\url{https://link.aps.org/doi/10.1103/RevModPhys.18.384}.

\bibitem[{\citenamefont{Piacentini et~al.}(1976)\citenamefont{Piacentini,
  Lynch, and Olson}}]{Piacentini}
\bibinfo{author}{\bibfnamefont{M.}~\bibnamefont{Piacentini}},
  \bibinfo{author}{\bibfnamefont{D.~W.} \bibnamefont{Lynch}}, \bibnamefont{and}
  \bibinfo{author}{\bibfnamefont{C.~G.} \bibnamefont{Olson}},
  \bibinfo{journal}{Phys. Rev. B} \textbf{\bibinfo{volume}{13}},
  \bibinfo{pages}{5530} (\bibinfo{year}{1976}),
  \urlprefix\url{https://link.aps.org/doi/10.1103/PhysRevB.13.5530}.

\bibitem[{\citenamefont{Nichelatti et~al.}(2019)\citenamefont{Nichelatti,
  Piccinini, Ampollini, Picardi, Ronsivalle, Bonfigli, Vincenti, and
  Montereali}}]{Nichelatti}
\bibinfo{author}{\bibfnamefont{E.}~\bibnamefont{Nichelatti}},
  \bibinfo{author}{\bibfnamefont{M.}~\bibnamefont{Piccinini}},
  \bibinfo{author}{\bibfnamefont{A.}~\bibnamefont{Ampollini}},
  \bibinfo{author}{\bibfnamefont{L.}~\bibnamefont{Picardi}},
  \bibinfo{author}{\bibfnamefont{C.}~\bibnamefont{Ronsivalle}},
  \bibinfo{author}{\bibfnamefont{F.}~\bibnamefont{Bonfigli}},
  \bibinfo{author}{\bibfnamefont{M.}~\bibnamefont{Vincenti}}, \bibnamefont{and}
  \bibinfo{author}{\bibfnamefont{R.}~\bibnamefont{Montereali}},
  \bibinfo{journal}{Optical Materials} \textbf{\bibinfo{volume}{89}},
  \bibinfo{pages}{414} (\bibinfo{year}{2019}), ISSN \bibinfo{issn}{0925-3467},
  \urlprefix\url{https://www.sciencedirect.com/science/article/pii/S0925346719300758}.

\bibitem[{\citenamefont{Baldacchini
  et~al.}(1996{\natexlab{a}})\citenamefont{Baldacchini, Cremona, d'Auria,
  Montereali, and Kalinov}}]{Baldacchini}
\bibinfo{author}{\bibfnamefont{G.}~\bibnamefont{Baldacchini}},
  \bibinfo{author}{\bibfnamefont{M.}~\bibnamefont{Cremona}},
  \bibinfo{author}{\bibfnamefont{G.}~\bibnamefont{d'Auria}},
  \bibinfo{author}{\bibfnamefont{R.~M.} \bibnamefont{Montereali}},
  \bibnamefont{and} \bibinfo{author}{\bibfnamefont{V.}~\bibnamefont{Kalinov}},
  \bibinfo{journal}{Phys. Rev. B} \textbf{\bibinfo{volume}{54}},
  \bibinfo{pages}{17508} (\bibinfo{year}{1996}{\natexlab{a}}),
  \urlprefix\url{https://link.aps.org/doi/10.1103/PhysRevB.54.17508}.

\bibitem[{\citenamefont{Ter-Mikirtychev and Tsuboi}(1996)}]{Ter-Mikirtychev}
\bibinfo{author}{\bibfnamefont{V.~V.} \bibnamefont{Ter-Mikirtychev}}
  \bibnamefont{and} \bibinfo{author}{\bibfnamefont{T.}~\bibnamefont{Tsuboi}},
  \bibinfo{journal}{Progress in Quantum Electronics}
  \textbf{\bibinfo{volume}{20}}, \bibinfo{pages}{219} (\bibinfo{year}{1996}),
  ISSN \bibinfo{issn}{0079-6727},
  \urlprefix\url{https://www.sciencedirect.com/science/article/pii/0079672796000018}.

\bibitem[{\citenamefont{Montereali et~al.}(2022)\citenamefont{Montereali,
  Bonfigli, Nichelatti, Nigro, Piccinini, and Vincenti}}]{Montereali}
\bibinfo{author}{\bibfnamefont{R.}~\bibnamefont{Montereali}},
  \bibinfo{author}{\bibfnamefont{F.}~\bibnamefont{Bonfigli}},
  \bibinfo{author}{\bibfnamefont{E.}~\bibnamefont{Nichelatti}},
  \bibinfo{author}{\bibfnamefont{V.}~\bibnamefont{Nigro}},
  \bibinfo{author}{\bibfnamefont{M.}~\bibnamefont{Piccinini}},
  \bibnamefont{and} \bibinfo{author}{\bibfnamefont{M.}~\bibnamefont{Vincenti}},
  \bibinfo{journal}{Journal of Physics: Conference Series}
  \textbf{\bibinfo{volume}{2298}}, \bibinfo{pages}{012001}
  (\bibinfo{year}{2022}),
  \urlprefix\url{https://dx.doi.org/10.1088/1742-6596/2298/1/012001}.

\bibitem[{\citenamefont{Gellermann}(1991)}]{Gellermann}
\bibinfo{author}{\bibfnamefont{W.}~\bibnamefont{Gellermann}},
  \bibinfo{journal}{Journal of Physics and Chemistry of Solids}
  \textbf{\bibinfo{volume}{52}}, \bibinfo{pages}{249} (\bibinfo{year}{1991}),
  ISSN \bibinfo{issn}{0022-3697},
  \urlprefix\url{https://www.sciencedirect.com/science/article/pii/002236979190068B}.

\bibitem[{\citenamefont{Montereali et~al.}(2017)\citenamefont{Montereali,
  Ampollini, Picardi, Ronsivalle, Bonfigli, Libera, Nichelatti, Piccinini, and
  Vincenti}}]{Montereali2}
\bibinfo{author}{\bibfnamefont{R.}~\bibnamefont{Montereali}},
  \bibinfo{author}{\bibfnamefont{A.}~\bibnamefont{Ampollini}},
  \bibinfo{author}{\bibfnamefont{L.}~\bibnamefont{Picardi}},
  \bibinfo{author}{\bibfnamefont{C.}~\bibnamefont{Ronsivalle}},
  \bibinfo{author}{\bibfnamefont{F.}~\bibnamefont{Bonfigli}},
  \bibinfo{author}{\bibfnamefont{S.}~\bibnamefont{Libera}},
  \bibinfo{author}{\bibfnamefont{E.}~\bibnamefont{Nichelatti}},
  \bibinfo{author}{\bibfnamefont{M.}~\bibnamefont{Piccinini}},
  \bibnamefont{and} \bibinfo{author}{\bibfnamefont{M.}~\bibnamefont{Vincenti}},
  \bibinfo{journal}{IOP Conference Series: Materials Science and Engineering}
  \textbf{\bibinfo{volume}{169}}, \bibinfo{pages}{012012}
  (\bibinfo{year}{2017}),
  \urlprefix\url{https://dx.doi.org/10.1088/1757-899X/169/1/012012}.

\bibitem[{\citenamefont{Cogswell et~al.}(2021)\citenamefont{Cogswell, Goel, and
  Huber}}]{PALEO}
\bibinfo{author}{\bibfnamefont{B.~K.} \bibnamefont{Cogswell}},
  \bibinfo{author}{\bibfnamefont{A.}~\bibnamefont{Goel}}, \bibnamefont{and}
  \bibinfo{author}{\bibfnamefont{P.}~\bibnamefont{Huber}},
  \bibinfo{journal}{Phys. Rev. Appl.} \textbf{\bibinfo{volume}{16}},
  \bibinfo{pages}{064060} (\bibinfo{year}{2021}),
  \urlprefix\url{https://link.aps.org/doi/10.1103/PhysRevApplied.16.064060}.

\bibitem[{\citenamefont{Ottmer}(1928)}]{Ottmer}
\bibinfo{author}{\bibfnamefont{R.}~\bibnamefont{Ottmer}},
  \bibinfo{journal}{Zeitschrift f{\"u}r Physik} \textbf{\bibinfo{volume}{46}},
  \bibinfo{pages}{798} (\bibinfo{year}{1928}), ISSN \bibinfo{issn}{0044-3328},
  \urlprefix\url{https://doi.org/10.1007/BF01391017}.

\bibitem[{\citenamefont{Pringsheim and Yuster}(1950)}]{Pringsheim}
\bibinfo{author}{\bibfnamefont{P.}~\bibnamefont{Pringsheim}} \bibnamefont{and}
  \bibinfo{author}{\bibfnamefont{P.}~\bibnamefont{Yuster}},
  \bibinfo{journal}{Phys. Rev.} \textbf{\bibinfo{volume}{78}},
  \bibinfo{pages}{293} (\bibinfo{year}{1950}),
  \urlprefix\url{https://link.aps.org/doi/10.1103/PhysRev.78.293}.

\bibitem[{\citenamefont{Klick}(1950)}]{Klick}
\bibinfo{author}{\bibfnamefont{C.~C.} \bibnamefont{Klick}},
  \bibinfo{journal}{Phys. Rev.} \textbf{\bibinfo{volume}{79}},
  \bibinfo{pages}{894} (\bibinfo{year}{1950}),
  \urlprefix\url{https://link.aps.org/doi/10.1103/PhysRev.79.894.2}.

\bibitem[{\citenamefont{Uchida and Yagi}(1952)}]{Uchida}
\bibinfo{author}{\bibfnamefont{Y.}~\bibnamefont{Uchida}} \bibnamefont{and}
  \bibinfo{author}{\bibfnamefont{H.}~\bibnamefont{Yagi}},
  \bibinfo{journal}{Journal of the Physical Society of Japan}
  \textbf{\bibinfo{volume}{7}}, \bibinfo{pages}{109} (\bibinfo{year}{1952}),
  \eprint{https://doi.org/10.1143/JPSJ.7.109},
  \urlprefix\url{https://doi.org/10.1143/JPSJ.7.109}.

\bibitem[{\citenamefont{Delbecq and Pringsheim}(1953)}]{Delbecq}
\bibinfo{author}{\bibfnamefont{C.~J.} \bibnamefont{Delbecq}} \bibnamefont{and}
  \bibinfo{author}{\bibfnamefont{P.}~\bibnamefont{Pringsheim}},
  \bibinfo{journal}{The Journal of Chemical Physics}
  \textbf{\bibinfo{volume}{21}}, \bibinfo{pages}{794} (\bibinfo{year}{1953}),
  ISSN \bibinfo{issn}{0021-9606},
  \eprint{https://pubs.aip.org/aip/jcp/article-pdf/21/5/794/18803101/794\_1\_online.pdf},
  \urlprefix\url{https://doi.org/10.1063/1.1699035}.

\bibitem[{\citenamefont{Bate and Heer}(1958)}]{Bate}
\bibinfo{author}{\bibfnamefont{R.}~\bibnamefont{Bate}} \bibnamefont{and}
  \bibinfo{author}{\bibfnamefont{C.}~\bibnamefont{Heer}},
  \bibinfo{journal}{Journal of Physics and Chemistry of Solids}
  \textbf{\bibinfo{volume}{7}}, \bibinfo{pages}{14} (\bibinfo{year}{1958}),
  ISSN \bibinfo{issn}{0022-3697},
  \urlprefix\url{https://www.sciencedirect.com/science/article/pii/0022369758901768}.

\bibitem[{\citenamefont{Gilman and Johnston}(1958)}]{Gilman}
\bibinfo{author}{\bibfnamefont{J.~J.} \bibnamefont{Gilman}} \bibnamefont{and}
  \bibinfo{author}{\bibfnamefont{W.~G.} \bibnamefont{Johnston}},
  \bibinfo{journal}{Journal of Applied Physics} \textbf{\bibinfo{volume}{29}},
  \bibinfo{pages}{877} (\bibinfo{year}{1958}), ISSN \bibinfo{issn}{0021-8979},
  \eprint{https://pubs.aip.org/aip/jap/article-pdf/29/6/877/18318182/877\_1\_online.pdf},
  \urlprefix\url{https://doi.org/10.1063/1.1723322}.

\bibitem[{\citenamefont{Kato et~al.}(1960)\citenamefont{Kato, Nakashima,
  Nakamura, and Uchida}}]{Kalo}
\bibinfo{author}{\bibfnamefont{R.}~\bibnamefont{Kato}},
  \bibinfo{author}{\bibfnamefont{S.-i.} \bibnamefont{Nakashima}},
  \bibinfo{author}{\bibfnamefont{K.}~\bibnamefont{Nakamura}}, \bibnamefont{and}
  \bibinfo{author}{\bibfnamefont{Y.}~\bibnamefont{Uchida}},
  \bibinfo{journal}{Journal of the Physical Society of Japan}
  \textbf{\bibinfo{volume}{15}}, \bibinfo{pages}{2111} (\bibinfo{year}{1960}),
  \eprint{https://doi.org/10.1143/JPSJ.15.2111},
  \urlprefix\url{https://doi.org/10.1143/JPSJ.15.2111}.

\bibitem[{\citenamefont{Wiegand and Smoluchowski}(1959)}]{Weigand}
\bibinfo{author}{\bibfnamefont{D.~A.} \bibnamefont{Wiegand}} \bibnamefont{and}
  \bibinfo{author}{\bibfnamefont{R.}~\bibnamefont{Smoluchowski}},
  \bibinfo{journal}{Phys. Rev.} \textbf{\bibinfo{volume}{116}},
  \bibinfo{pages}{1069} (\bibinfo{year}{1959}),
  \urlprefix\url{https://link.aps.org/doi/10.1103/PhysRev.116.1069}.

\bibitem[{\citenamefont{Fischer}(1959)}]{Fischer}
\bibinfo{author}{\bibfnamefont{F.}~\bibnamefont{Fischer}},
  \bibinfo{journal}{Zeitschrift f{\"u}r Physik} \textbf{\bibinfo{volume}{154}},
  \bibinfo{pages}{534} (\bibinfo{year}{1959}), ISSN \bibinfo{issn}{0044-3328},
  \urlprefix\url{https://doi.org/10.1007/BF01337564}.

\bibitem[{\citenamefont{Mott and Gurney}(1948)}]{ionic_book}
\bibinfo{author}{\bibfnamefont{N.~F.} \bibnamefont{Mott}} \bibnamefont{and}
  \bibinfo{author}{\bibfnamefont{R.~W.} \bibnamefont{Gurney}},
  \emph{\bibinfo{title}{Electronic Processes in Ionic Crystals}}
  (\bibinfo{publisher}{Oxford University Press}, \bibinfo{year}{1948}).

\bibitem[{\citenamefont{Kaufman and Clark}(1963)}]{Kaufman}
\bibinfo{author}{\bibfnamefont{J.~V.~R.} \bibnamefont{Kaufman}}
  \bibnamefont{and} \bibinfo{author}{\bibfnamefont{C.~D.} \bibnamefont{Clark}},
  \bibinfo{journal}{The Journal of Chemical Physics}
  \textbf{\bibinfo{volume}{38}}, \bibinfo{pages}{1388} (\bibinfo{year}{1963}),
  ISSN \bibinfo{issn}{0021-9606},
  \eprint{https://pubs.aip.org/aip/jcp/article-pdf/38/6/1388/18828455/1388\_1\_online.pdf},
  \urlprefix\url{https://doi.org/10.1063/1.1733863}.

\bibitem[{\citenamefont{Görlich et~al.}(1963)\citenamefont{Görlich, Karras,
  and Kötitz}}]{Gorlich}
\bibinfo{author}{\bibfnamefont{P.}~\bibnamefont{Görlich}},
  \bibinfo{author}{\bibfnamefont{H.}~\bibnamefont{Karras}}, \bibnamefont{and}
  \bibinfo{author}{\bibfnamefont{G.}~\bibnamefont{Kötitz}},
  \bibinfo{journal}{physica status solidi (b)} \textbf{\bibinfo{volume}{3}},
  \bibinfo{pages}{1629} (\bibinfo{year}{1963}),
  \eprint{https://onlinelibrary.wiley.com/doi/pdf/10.1002/pssb.19630030909},
  \urlprefix\url{https://onlinelibrary.wiley.com/doi/abs/10.1002/pssb.19630030909}.

\bibitem[{\citenamefont{Nahum and Wiegand}(1967)}]{Nahum}
\bibinfo{author}{\bibfnamefont{J.}~\bibnamefont{Nahum}} \bibnamefont{and}
  \bibinfo{author}{\bibfnamefont{D.~A.} \bibnamefont{Wiegand}},
  \bibinfo{journal}{Phys. Rev.} \textbf{\bibinfo{volume}{154}},
  \bibinfo{pages}{817} (\bibinfo{year}{1967}),
  \urlprefix\url{https://link.aps.org/doi/10.1103/PhysRev.154.817}.

\bibitem[{\citenamefont{Norman et~al.}(2021)\citenamefont{Norman, Majety, Wang,
  Casey, Curro, and Radulaski}}]{Norman}
\bibinfo{author}{\bibfnamefont{V.~A.} \bibnamefont{Norman}},
  \bibinfo{author}{\bibfnamefont{S.}~\bibnamefont{Majety}},
  \bibinfo{author}{\bibfnamefont{Z.}~\bibnamefont{Wang}},
  \bibinfo{author}{\bibfnamefont{W.~H.} \bibnamefont{Casey}},
  \bibinfo{author}{\bibfnamefont{N.}~\bibnamefont{Curro}}, \bibnamefont{and}
  \bibinfo{author}{\bibfnamefont{M.}~\bibnamefont{Radulaski}},
  \bibinfo{journal}{InfoMat} \textbf{\bibinfo{volume}{3}}, \bibinfo{pages}{869}
  (\bibinfo{year}{2021}),
  \eprint{https://onlinelibrary.wiley.com/doi/pdf/10.1002/inf2.12128},
  \urlprefix\url{https://onlinelibrary.wiley.com/doi/abs/10.1002/inf2.12128}.

\bibitem[{\citenamefont{Wasielewski et~al.}(2020)\citenamefont{Wasielewski,
  Forbes, Frank, Kowalski, Scholes, Yuen-Zhou, Baldo, Freedman, Goldsmith,
  Goodson et~al.}}]{Wasielewski}
\bibinfo{author}{\bibfnamefont{M.~R.} \bibnamefont{Wasielewski}},
  \bibinfo{author}{\bibfnamefont{M.~D.~E.} \bibnamefont{Forbes}},
  \bibinfo{author}{\bibfnamefont{N.~L.} \bibnamefont{Frank}},
  \bibinfo{author}{\bibfnamefont{K.}~\bibnamefont{Kowalski}},
  \bibinfo{author}{\bibfnamefont{G.~D.} \bibnamefont{Scholes}},
  \bibinfo{author}{\bibfnamefont{J.}~\bibnamefont{Yuen-Zhou}},
  \bibinfo{author}{\bibfnamefont{M.~A.} \bibnamefont{Baldo}},
  \bibinfo{author}{\bibfnamefont{D.~E.} \bibnamefont{Freedman}},
  \bibinfo{author}{\bibfnamefont{R.~H.} \bibnamefont{Goldsmith}},
  \bibinfo{author}{\bibfnamefont{T.}~\bibnamefont{Goodson}},
  \bibnamefont{et~al.}, \bibinfo{journal}{Nature Reviews Chemistry}
  \textbf{\bibinfo{volume}{4}}, \bibinfo{pages}{490} (\bibinfo{year}{2020}),
  ISSN \bibinfo{issn}{2397-3358},
  \urlprefix\url{https://doi.org/10.1038/s41570-020-0200-5}.

\bibitem[{\citenamefont{Bayliss et~al.}(2020)\citenamefont{Bayliss, Laorenza,
  Mintun, Kovos, Freedman, and Awschalom}}]{Bayliss}
\bibinfo{author}{\bibfnamefont{S.~L.} \bibnamefont{Bayliss}},
  \bibinfo{author}{\bibfnamefont{D.~W.} \bibnamefont{Laorenza}},
  \bibinfo{author}{\bibfnamefont{P.~J.} \bibnamefont{Mintun}},
  \bibinfo{author}{\bibfnamefont{B.~D.} \bibnamefont{Kovos}},
  \bibinfo{author}{\bibfnamefont{D.~E.} \bibnamefont{Freedman}},
  \bibnamefont{and} \bibinfo{author}{\bibfnamefont{D.~D.}
  \bibnamefont{Awschalom}}, \bibinfo{journal}{Science}
  \textbf{\bibinfo{volume}{370}}, \bibinfo{pages}{1309} (\bibinfo{year}{2020}),
  \eprint{https://www.science.org/doi/pdf/10.1126/science.abb9352},
  \urlprefix\url{https://www.science.org/doi/abs/10.1126/science.abb9352}.

\bibitem[{\citenamefont{Duarte et~al.}(1994{\natexlab{a}})\citenamefont{Duarte,
  Ranieri, and Vieira}}]{Duarte}
\bibinfo{author}{\bibfnamefont{M.}~\bibnamefont{Duarte}},
  \bibinfo{author}{\bibfnamefont{I.~M.} \bibnamefont{Ranieri}},
  \bibnamefont{and} \bibinfo{author}{\bibfnamefont{M.~M.}
  \bibnamefont{Vieira}}, \bibinfo{journal}{Optical Materials}
  \textbf{\bibinfo{volume}{3}}, \bibinfo{pages}{269}
  (\bibinfo{year}{1994}{\natexlab{a}}),
  \urlprefix\url{https://doi.org/10.1016/0925-3467(94)90040-x}.

\bibitem[{\citenamefont{Ivanov et~al.}(2023)\citenamefont{Ivanov, Ivanov,
  Simoni, Parajuli, Kanté, Schenkel, and Tan}}]{Ivanov-db}
\bibinfo{author}{\bibfnamefont{V.}~\bibnamefont{Ivanov}},
  \bibinfo{author}{\bibfnamefont{A.}~\bibnamefont{Ivanov}},
  \bibinfo{author}{\bibfnamefont{J.}~\bibnamefont{Simoni}},
  \bibinfo{author}{\bibfnamefont{P.}~\bibnamefont{Parajuli}},
  \bibinfo{author}{\bibfnamefont{B.}~\bibnamefont{Kanté}},
  \bibinfo{author}{\bibfnamefont{T.}~\bibnamefont{Schenkel}}, \bibnamefont{and}
  \bibinfo{author}{\bibfnamefont{L.}~\bibnamefont{Tan}},
  \emph{\bibinfo{title}{Database of semiconductor point-defect properties for
  applications in quantum technologies}} (\bibinfo{year}{2023}),
  \eprint{2303.16283}, \urlprefix\url{https://arxiv.org/abs/2303.16283}.

\bibitem[{\citenamefont{Xiong et~al.}(2023)\citenamefont{Xiong, Bourgois,
  Sheremetyeva, Chen, Dahliah, Song, Zheng, Griffin, Sipahigil, and
  Hautier}}]{xiong-db}
\bibinfo{author}{\bibfnamefont{Y.}~\bibnamefont{Xiong}},
  \bibinfo{author}{\bibfnamefont{C.}~\bibnamefont{Bourgois}},
  \bibinfo{author}{\bibfnamefont{N.}~\bibnamefont{Sheremetyeva}},
  \bibinfo{author}{\bibfnamefont{W.}~\bibnamefont{Chen}},
  \bibinfo{author}{\bibfnamefont{D.}~\bibnamefont{Dahliah}},
  \bibinfo{author}{\bibfnamefont{H.}~\bibnamefont{Song}},
  \bibinfo{author}{\bibfnamefont{J.}~\bibnamefont{Zheng}},
  \bibinfo{author}{\bibfnamefont{S.~M.} \bibnamefont{Griffin}},
  \bibinfo{author}{\bibfnamefont{A.}~\bibnamefont{Sipahigil}},
  \bibnamefont{and} \bibinfo{author}{\bibfnamefont{G.}~\bibnamefont{Hautier}},
  \bibinfo{journal}{Science Advances} \textbf{\bibinfo{volume}{9}},
  \bibinfo{pages}{eadh8617} (\bibinfo{year}{2023}),
  \eprint{https://www.science.org/doi/pdf/10.1126/sciadv.adh8617},
  \urlprefix\url{https://www.science.org/doi/abs/10.1126/sciadv.adh8617}.

\bibitem[{\citenamefont{Davidsson et~al.}(2021)\citenamefont{Davidsson, Ivády,
  Armiento, and Abrikosov}}]{davidsson-db}
\bibinfo{author}{\bibfnamefont{J.}~\bibnamefont{Davidsson}},
  \bibinfo{author}{\bibfnamefont{V.}~\bibnamefont{Ivády}},
  \bibinfo{author}{\bibfnamefont{R.}~\bibnamefont{Armiento}}, \bibnamefont{and}
  \bibinfo{author}{\bibfnamefont{I.~A.} \bibnamefont{Abrikosov}},
  \bibinfo{journal}{Computer Physics Communications}
  \textbf{\bibinfo{volume}{269}}, \bibinfo{pages}{108091}
  (\bibinfo{year}{2021}), ISSN \bibinfo{issn}{0010-4655},
  \urlprefix\url{https://www.sciencedirect.com/science/article/pii/S0010465521002034}.

\bibitem[{\citenamefont{Heyd and Scuseria}(2004)}]{HSE}
\bibinfo{author}{\bibfnamefont{J.}~\bibnamefont{Heyd}} \bibnamefont{and}
  \bibinfo{author}{\bibfnamefont{G.~E.} \bibnamefont{Scuseria}},
  \bibinfo{journal}{The Journal of Chemical Physics}
  \textbf{\bibinfo{volume}{121}}, \bibinfo{pages}{1187} (\bibinfo{year}{2004}),
  ISSN \bibinfo{issn}{0021-9606},
  \eprint{https://pubs.aip.org/aip/jcp/article-pdf/121/3/1187/19171502/1187\_1\_online.pdf},
  \urlprefix\url{https://doi.org/10.1063/1.1760074}.

\bibitem[{\citenamefont{Duerig and Markham}(1952)}]{Duerig}
\bibinfo{author}{\bibfnamefont{W.~H.} \bibnamefont{Duerig}} \bibnamefont{and}
  \bibinfo{author}{\bibfnamefont{J.~J.} \bibnamefont{Markham}},
  \bibinfo{journal}{Phys. Rev.} \textbf{\bibinfo{volume}{88}},
  \bibinfo{pages}{1043} (\bibinfo{year}{1952}),
  \urlprefix\url{https://link.aps.org/doi/10.1103/PhysRev.88.1043}.

\bibitem[{\citenamefont{Petroff}(1950)}]{Petroff}
\bibinfo{author}{\bibfnamefont{S.}~\bibnamefont{Petroff}},
  \bibinfo{journal}{Zeitschrift f{\"u}r Physik} \textbf{\bibinfo{volume}{127}},
  \bibinfo{pages}{443} (\bibinfo{year}{1950}), ISSN \bibinfo{issn}{0044-3328},
  \urlprefix\url{https://doi.org/10.1007/BF01329841}.

\bibitem[{\citenamefont{Ivey}(1947)}]{Ivey}
\bibinfo{author}{\bibfnamefont{H.~F.} \bibnamefont{Ivey}},
  \bibinfo{journal}{Phys. Rev.} \textbf{\bibinfo{volume}{72}},
  \bibinfo{pages}{341} (\bibinfo{year}{1947}),
  \urlprefix\url{https://link.aps.org/doi/10.1103/PhysRev.72.341}.

\bibitem[{\citenamefont{van Doorn}(1960)}]{vanDoorn}
\bibinfo{author}{\bibfnamefont{C.~Z.} \bibnamefont{van Doorn}},
  \bibinfo{journal}{Phys. Rev. Lett.} \textbf{\bibinfo{volume}{4}},
  \bibinfo{pages}{236} (\bibinfo{year}{1960}),
  \urlprefix\url{https://link.aps.org/doi/10.1103/PhysRevLett.4.236}.

\bibitem[{\citenamefont{Faraday
  et~al.}(1961{\natexlab{a}})\citenamefont{Faraday, Rabin, and
  Compton}}]{Faraday}
\bibinfo{author}{\bibfnamefont{B.~J.} \bibnamefont{Faraday}},
  \bibinfo{author}{\bibfnamefont{H.}~\bibnamefont{Rabin}}, \bibnamefont{and}
  \bibinfo{author}{\bibfnamefont{W.~D.} \bibnamefont{Compton}},
  \bibinfo{journal}{Phys. Rev. Lett.} \textbf{\bibinfo{volume}{7}},
  \bibinfo{pages}{433} (\bibinfo{year}{1961}{\natexlab{a}}),
  \urlprefix\url{https://link.aps.org/doi/10.1103/PhysRevLett.7.433}.

\bibitem[{\citenamefont{Okuda}(1961{\natexlab{a}})}]{Okuda}
\bibinfo{author}{\bibfnamefont{A.}~\bibnamefont{Okuda}},
  \bibinfo{journal}{Journal of the Physical Society of Japan}
  \textbf{\bibinfo{volume}{16}}, \bibinfo{pages}{1746}
  (\bibinfo{year}{1961}{\natexlab{a}}),
  \eprint{https://doi.org/10.1143/JPSJ.16.1746},
  \urlprefix\url{https://doi.org/10.1143/JPSJ.16.1746}.

\bibitem[{\citenamefont{Baldacchini}(2002)}]{Baldacchini_LiF_seasons}
\bibinfo{author}{\bibfnamefont{G.}~\bibnamefont{Baldacchini}},
  \bibinfo{journal}{Journal of Luminescence} \textbf{\bibinfo{volume}{100}},
  \bibinfo{pages}{333} (\bibinfo{year}{2002}), ISSN \bibinfo{issn}{0022-2313},
  \urlprefix\url{https://www.sciencedirect.com/science/article/pii/S002223130200460X}.

\bibitem[{\citenamefont{Seitz}(1954)}]{Seitz_1954}
\bibinfo{author}{\bibfnamefont{F.}~\bibnamefont{Seitz}}, \bibinfo{journal}{Rev.
  Mod. Phys.} \textbf{\bibinfo{volume}{26}}, \bibinfo{pages}{7}
  (\bibinfo{year}{1954}),
  \urlprefix\url{https://link.aps.org/doi/10.1103/RevModPhys.26.7}.

\bibitem[{\citenamefont{Chaney}(1976)}]{Chaney_LCAO}
\bibinfo{author}{\bibfnamefont{R.~C.} \bibnamefont{Chaney}},
  \bibinfo{journal}{Phys. Rev. B} \textbf{\bibinfo{volume}{14}},
  \bibinfo{pages}{4578} (\bibinfo{year}{1976}),
  \urlprefix\url{https://link.aps.org/doi/10.1103/PhysRevB.14.4578}.

\bibitem[{\citenamefont{Pederson and Klein}(1988)}]{Pederson}
\bibinfo{author}{\bibfnamefont{M.~R.} \bibnamefont{Pederson}} \bibnamefont{and}
  \bibinfo{author}{\bibfnamefont{B.~M.} \bibnamefont{Klein}},
  \bibinfo{journal}{Phys. Rev. B} \textbf{\bibinfo{volume}{37}},
  \bibinfo{pages}{10319} (\bibinfo{year}{1988}),
  \urlprefix\url{https://link.aps.org/doi/10.1103/PhysRevB.37.10319}.

\bibitem[{\citenamefont{Mallia et~al.}(2001)\citenamefont{Mallia, Orlando,
  Roetti, Ugliengo, and Dovesi}}]{Mallia}
\bibinfo{author}{\bibfnamefont{G.}~\bibnamefont{Mallia}},
  \bibinfo{author}{\bibfnamefont{R.}~\bibnamefont{Orlando}},
  \bibinfo{author}{\bibfnamefont{C.}~\bibnamefont{Roetti}},
  \bibinfo{author}{\bibfnamefont{P.}~\bibnamefont{Ugliengo}}, \bibnamefont{and}
  \bibinfo{author}{\bibfnamefont{R.}~\bibnamefont{Dovesi}},
  \bibinfo{journal}{Phys. Rev. B} \textbf{\bibinfo{volume}{63}},
  \bibinfo{pages}{235102} (\bibinfo{year}{2001}),
  \urlprefix\url{https://link.aps.org/doi/10.1103/PhysRevB.63.235102}.

\bibitem[{\citenamefont{K{\"o}lmel and
  Ewig}(2001)}]{Komel_F_M_embedded_cluster}
\bibinfo{author}{\bibfnamefont{C.}~\bibnamefont{K{\"o}lmel}} \bibnamefont{and}
  \bibinfo{author}{\bibfnamefont{C.~S.} \bibnamefont{Ewig}},
  \bibinfo{journal}{The Journal of Physical Chemistry B}
  \textbf{\bibinfo{volume}{105}}, \bibinfo{pages}{8538} (\bibinfo{year}{2001}),
  \eprint{https://doi.org/10.1021/jp012155e},
  \urlprefix\url{https://doi.org/10.1021/jp012155e}.

\bibitem[{\citenamefont{Ewig et~al.}(1992)\citenamefont{Ewig, Tellinghuisen,
  and Mendenhall}}]{Ewig_cluster}
\bibinfo{author}{\bibfnamefont{C.~S.} \bibnamefont{Ewig}},
  \bibinfo{author}{\bibfnamefont{J.}~\bibnamefont{Tellinghuisen}},
  \bibnamefont{and} \bibinfo{author}{\bibfnamefont{M.~H.}
  \bibnamefont{Mendenhall}}, \bibinfo{journal}{Chemical Physics Letters}
  \textbf{\bibinfo{volume}{188}}, \bibinfo{pages}{501} (\bibinfo{year}{1992}),
  ISSN \bibinfo{issn}{0009-2614},
  \urlprefix\url{https://www.sciencedirect.com/science/article/pii/0009261492808567}.

\bibitem[{\citenamefont{Karsai et~al.}(2014)\citenamefont{Karsai, Tiwald,
  Laskowski, Tran, Koller, Gr\"afe, Burgd\"orfer, Wirtz, and
  Blaha}}]{Blaha_Fcenter}
\bibinfo{author}{\bibfnamefont{F.}~\bibnamefont{Karsai}},
  \bibinfo{author}{\bibfnamefont{P.}~\bibnamefont{Tiwald}},
  \bibinfo{author}{\bibfnamefont{R.}~\bibnamefont{Laskowski}},
  \bibinfo{author}{\bibfnamefont{F.}~\bibnamefont{Tran}},
  \bibinfo{author}{\bibfnamefont{D.}~\bibnamefont{Koller}},
  \bibinfo{author}{\bibfnamefont{S.}~\bibnamefont{Gr\"afe}},
  \bibinfo{author}{\bibfnamefont{J.}~\bibnamefont{Burgd\"orfer}},
  \bibinfo{author}{\bibfnamefont{L.}~\bibnamefont{Wirtz}}, \bibnamefont{and}
  \bibinfo{author}{\bibfnamefont{P.}~\bibnamefont{Blaha}},
  \bibinfo{journal}{Phys. Rev. B} \textbf{\bibinfo{volume}{89}},
  \bibinfo{pages}{125429} (\bibinfo{year}{2014}),
  \urlprefix\url{https://link.aps.org/doi/10.1103/PhysRevB.89.125429}.

\bibitem[{\citenamefont{Wang and Klein}(1981{\natexlab{a}})}]{DFT-gaps}
\bibinfo{author}{\bibfnamefont{C.~S.} \bibnamefont{Wang}} \bibnamefont{and}
  \bibinfo{author}{\bibfnamefont{B.~M.} \bibnamefont{Klein}},
  \bibinfo{journal}{Phys. Rev. B} \textbf{\bibinfo{volume}{24}},
  \bibinfo{pages}{3393} (\bibinfo{year}{1981}{\natexlab{a}}),
  \urlprefix\url{https://link.aps.org/doi/10.1103/PhysRevB.24.3393}.

\bibitem[{\citenamefont{Chen and Pasquarello}(2013)}]{Chen}
\bibinfo{author}{\bibfnamefont{W.}~\bibnamefont{Chen}} \bibnamefont{and}
  \bibinfo{author}{\bibfnamefont{A.}~\bibnamefont{Pasquarello}},
  \bibinfo{journal}{Phys. Rev. B} \textbf{\bibinfo{volume}{88}},
  \bibinfo{pages}{115104} (\bibinfo{year}{2013}),
  \urlprefix\url{https://link.aps.org/doi/10.1103/PhysRevB.88.115104}.

\bibitem[{\citenamefont{Kresse and Joubert}(1999)}]{Kresse_potentials}
\bibinfo{author}{\bibfnamefont{G.}~\bibnamefont{Kresse}} \bibnamefont{and}
  \bibinfo{author}{\bibfnamefont{D.}~\bibnamefont{Joubert}},
  \bibinfo{journal}{Phys. Rev. B} \textbf{\bibinfo{volume}{59}},
  \bibinfo{pages}{1758} (\bibinfo{year}{1999}),
  \urlprefix\url{https://link.aps.org/doi/10.1103/PhysRevB.59.1758}.

\bibitem[{\citenamefont{Momma and Izumi}(2011)}]{Momma}
\bibinfo{author}{\bibfnamefont{K.}~\bibnamefont{Momma}} \bibnamefont{and}
  \bibinfo{author}{\bibfnamefont{F.}~\bibnamefont{Izumi}},
  \bibinfo{journal}{Journal of Applied Crystallography}
  \textbf{\bibinfo{volume}{44}}, \bibinfo{pages}{1272} (\bibinfo{year}{2011}),
  \urlprefix\url{https://doi.org/10.1107/S0021889811038970}.

\bibitem[{\citenamefont{Wang et~al.}(2021)\citenamefont{Wang, Xu, Liu, Tang,
  and Geng}}]{VASPKIT}
\bibinfo{author}{\bibfnamefont{V.}~\bibnamefont{Wang}},
  \bibinfo{author}{\bibfnamefont{N.}~\bibnamefont{Xu}},
  \bibinfo{author}{\bibfnamefont{J.-C.} \bibnamefont{Liu}},
  \bibinfo{author}{\bibfnamefont{G.}~\bibnamefont{Tang}}, \bibnamefont{and}
  \bibinfo{author}{\bibfnamefont{W.-T.} \bibnamefont{Geng}},
  \bibinfo{journal}{Computer Physics Communications}
  \textbf{\bibinfo{volume}{267}}, \bibinfo{pages}{108033}
  (\bibinfo{year}{2021}), ISSN \bibinfo{issn}{0010-4655},
  \urlprefix\url{https://www.sciencedirect.com/science/article/pii/S0010465521001454}.

\bibitem[{\citenamefont{Meyer and Wood}(1964)}]{Meyer_F2}
\bibinfo{author}{\bibfnamefont{A.}~\bibnamefont{Meyer}} \bibnamefont{and}
  \bibinfo{author}{\bibfnamefont{R.~F.} \bibnamefont{Wood}},
  \bibinfo{journal}{Phys. Rev.} \textbf{\bibinfo{volume}{133}},
  \bibinfo{pages}{A1436} (\bibinfo{year}{1964}),
  \urlprefix\url{https://link.aps.org/doi/10.1103/PhysRev.133.A1436}.

\bibitem[{\citenamefont{Faraday
  et~al.}(1961{\natexlab{b}})\citenamefont{Faraday, Rabin, and
  Compton}}]{Faraday_F2}
\bibinfo{author}{\bibfnamefont{B.~J.} \bibnamefont{Faraday}},
  \bibinfo{author}{\bibfnamefont{H.}~\bibnamefont{Rabin}}, \bibnamefont{and}
  \bibinfo{author}{\bibfnamefont{W.~D.} \bibnamefont{Compton}},
  \bibinfo{journal}{Phys. Rev. Lett.} \textbf{\bibinfo{volume}{7}},
  \bibinfo{pages}{57} (\bibinfo{year}{1961}{\natexlab{b}}),
  \urlprefix\url{https://link.aps.org/doi/10.1103/PhysRevLett.7.57}.

\bibitem[{\citenamefont{Tsuboi}(2000)}]{Tsuboi_F2_oscillation}
\bibinfo{author}{\bibfnamefont{T.}~\bibnamefont{Tsuboi}},
  \bibinfo{journal}{Nuclear Instruments and Methods in Physics Research Section
  B: Beam Interactions with Materials and Atoms}
  \textbf{\bibinfo{volume}{166-167}}, \bibinfo{pages}{804}
  (\bibinfo{year}{2000}), ISSN \bibinfo{issn}{0168-583X},
  \urlprefix\url{https://www.sciencedirect.com/science/article/pii/S0168583X99010605}.

\bibitem[{\citenamefont{Lisitsyna}(2007)}]{Lisitsyna_EP}
\bibinfo{author}{\bibfnamefont{L.}~\bibnamefont{Lisitsyna}},
  \bibinfo{journal}{physica status solidi c} \textbf{\bibinfo{volume}{4}},
  \bibinfo{pages}{1114} (\bibinfo{year}{2007}),
  \eprint{https://onlinelibrary.wiley.com/doi/pdf/10.1002/pssc.200673715},
  \urlprefix\url{https://onlinelibrary.wiley.com/doi/abs/10.1002/pssc.200673715}.

\bibitem[{\citenamefont{Lisitsyna}(1994)}]{Lisitsyna1994_F2_triplet}
\bibinfo{author}{\bibfnamefont{L.~A.} \bibnamefont{Lisitsyna}},
  \bibinfo{journal}{Russian Physics Journal} \textbf{\bibinfo{volume}{37}},
  \bibinfo{pages}{1024} (\bibinfo{year}{1994}), ISSN \bibinfo{issn}{1573-9228},
  \urlprefix\url{https://doi.org/10.1007/BF00559210}.

\bibitem[{\citenamefont{Okuda}(1961{\natexlab{b}})}]{Okuda_R111}
\bibinfo{author}{\bibfnamefont{A.}~\bibnamefont{Okuda}},
  \bibinfo{journal}{Journal of the Physical Society of Japan}
  \textbf{\bibinfo{volume}{16}}, \bibinfo{pages}{1746}
  (\bibinfo{year}{1961}{\natexlab{b}}),
  \eprint{https://doi.org/10.1143/JPSJ.16.1746},
  \urlprefix\url{https://doi.org/10.1143/JPSJ.16.1746}.

\bibitem[{\citenamefont{Duarte et~al.}(1994{\natexlab{b}})\citenamefont{Duarte,
  Ranieri, and Vieira}}]{Duarte_laser}
\bibinfo{author}{\bibfnamefont{M.}~\bibnamefont{Duarte}},
  \bibinfo{author}{\bibfnamefont{I.~M.} \bibnamefont{Ranieri}},
  \bibnamefont{and} \bibinfo{author}{\bibfnamefont{M.~M.}
  \bibnamefont{Vieira}}, \bibinfo{journal}{Optical Materials}
  \textbf{\bibinfo{volume}{3}}, \bibinfo{pages}{269}
  (\bibinfo{year}{1994}{\natexlab{b}}), ISSN \bibinfo{issn}{0925-3467},
  \urlprefix\url{https://www.sciencedirect.com/science/article/pii/092534679490040X}.

\bibitem[{\citenamefont{Tsuboi}(1999)}]{Tsuboi1999_F3m}
\bibinfo{author}{\bibfnamefont{T.}~\bibnamefont{Tsuboi}},
  \bibinfo{journal}{Applied Physics B} \textbf{\bibinfo{volume}{69}},
  \bibinfo{pages}{81} (\bibinfo{year}{1999}), ISSN \bibinfo{issn}{1432-0649},
  \urlprefix\url{https://doi.org/10.1007/s003400050773}.

\bibitem[{\citenamefont{Wang and Chu}(1966)}]{Wang_F3_states}
\bibinfo{author}{\bibfnamefont{S.-F.} \bibnamefont{Wang}} \bibnamefont{and}
  \bibinfo{author}{\bibfnamefont{C.}~\bibnamefont{Chu}},
  \bibinfo{journal}{Phys. Rev.} \textbf{\bibinfo{volume}{147}},
  \bibinfo{pages}{527} (\bibinfo{year}{1966}),
  \urlprefix\url{https://link.aps.org/doi/10.1103/PhysRev.147.527}.

\bibitem[{\citenamefont{Sati}(1976)}]{Sati_F3p}
\bibinfo{author}{\bibfnamefont{R.}~\bibnamefont{Sati}},
  \bibinfo{journal}{physica status solidi (b)} \textbf{\bibinfo{volume}{73}},
  \bibinfo{pages}{353} (\bibinfo{year}{1976}),
  \eprint{https://onlinelibrary.wiley.com/doi/pdf/10.1002/pssb.2220730135},
  \urlprefix\url{https://onlinelibrary.wiley.com/doi/abs/10.1002/pssb.2220730135}.

\bibitem[{\citenamefont{Baldacchini
  et~al.}(1996{\natexlab{b}})\citenamefont{Baldacchini, Cremona, d'Auria,
  Montereali, and Kalinov}}]{Baldacchini_F3p_triplet}
\bibinfo{author}{\bibfnamefont{G.}~\bibnamefont{Baldacchini}},
  \bibinfo{author}{\bibfnamefont{M.}~\bibnamefont{Cremona}},
  \bibinfo{author}{\bibfnamefont{G.}~\bibnamefont{d'Auria}},
  \bibinfo{author}{\bibfnamefont{R.~M.} \bibnamefont{Montereali}},
  \bibnamefont{and} \bibinfo{author}{\bibfnamefont{V.}~\bibnamefont{Kalinov}},
  \bibinfo{journal}{Phys. Rev. B} \textbf{\bibinfo{volume}{54}},
  \bibinfo{pages}{17508} (\bibinfo{year}{1996}{\natexlab{b}}),
  \urlprefix\url{https://link.aps.org/doi/10.1103/PhysRevB.54.17508}.

\bibitem[{sup()}]{supplement}
\bibinfo{note}{The Supplemental Material at [URL will be inserted by publisher]
  contains further details about the defect properties and calculations.}

\bibitem[{\citenamefont{Wang and Klein}(1981{\natexlab{b}})}]{DFT-gaps2}
\bibinfo{author}{\bibfnamefont{C.~S.} \bibnamefont{Wang}} \bibnamefont{and}
  \bibinfo{author}{\bibfnamefont{B.~M.} \bibnamefont{Klein}},
  \bibinfo{journal}{Phys. Rev. B} \textbf{\bibinfo{volume}{24}},
  \bibinfo{pages}{3417} (\bibinfo{year}{1981}{\natexlab{b}}),
  \urlprefix\url{https://link.aps.org/doi/10.1103/PhysRevB.24.3417}.

\bibitem[{\citenamefont{De\'ak et~al.}(2010)\citenamefont{De\'ak, Aradi,
  Frauenheim, Janz\'en, and Gali}}]{Deak}
\bibinfo{author}{\bibfnamefont{P.}~\bibnamefont{De\'ak}},
  \bibinfo{author}{\bibfnamefont{B.}~\bibnamefont{Aradi}},
  \bibinfo{author}{\bibfnamefont{T.}~\bibnamefont{Frauenheim}},
  \bibinfo{author}{\bibfnamefont{E.}~\bibnamefont{Janz\'en}}, \bibnamefont{and}
  \bibinfo{author}{\bibfnamefont{A.}~\bibnamefont{Gali}},
  \bibinfo{journal}{Phys. Rev. B} \textbf{\bibinfo{volume}{81}},
  \bibinfo{pages}{153203} (\bibinfo{year}{2010}),
  \urlprefix\url{https://link.aps.org/doi/10.1103/PhysRevB.81.153203}.

\bibitem[{\citenamefont{De\'ak et~al.}(2019)\citenamefont{De\'ak, Lorke, Aradi,
  and Frauenheim}}]{Deak2}
\bibinfo{author}{\bibfnamefont{P.}~\bibnamefont{De\'ak}},
  \bibinfo{author}{\bibfnamefont{M.}~\bibnamefont{Lorke}},
  \bibinfo{author}{\bibfnamefont{B.}~\bibnamefont{Aradi}}, \bibnamefont{and}
  \bibinfo{author}{\bibfnamefont{T.}~\bibnamefont{Frauenheim}},
  \bibinfo{journal}{Phys. Rev. B} \textbf{\bibinfo{volume}{99}},
  \bibinfo{pages}{085206} (\bibinfo{year}{2019}),
  \urlprefix\url{https://link.aps.org/doi/10.1103/PhysRevB.99.085206}.

\bibitem[{\citenamefont{Iv\'ady et~al.}(2013)\citenamefont{Iv\'ady, Abrikosov,
  Janz\'en, and Gali}}]{Ivady_screening}
\bibinfo{author}{\bibfnamefont{V.}~\bibnamefont{Iv\'ady}},
  \bibinfo{author}{\bibfnamefont{I.~A.} \bibnamefont{Abrikosov}},
  \bibinfo{author}{\bibfnamefont{E.}~\bibnamefont{Janz\'en}}, \bibnamefont{and}
  \bibinfo{author}{\bibfnamefont{A.}~\bibnamefont{Gali}},
  \bibinfo{journal}{Phys. Rev. B} \textbf{\bibinfo{volume}{87}},
  \bibinfo{pages}{205201} (\bibinfo{year}{2013}),
  \urlprefix\url{https://link.aps.org/doi/10.1103/PhysRevB.87.205201}.

\bibitem[{\citenamefont{Kresse and Hafner}(1993)}]{Kresse_GGA}
\bibinfo{author}{\bibfnamefont{G.}~\bibnamefont{Kresse}} \bibnamefont{and}
  \bibinfo{author}{\bibfnamefont{J.}~\bibnamefont{Hafner}},
  \bibinfo{journal}{Phys. Rev. B} \textbf{\bibinfo{volume}{47}},
  \bibinfo{pages}{558} (\bibinfo{year}{1993}),
  \urlprefix\url{https://link.aps.org/doi/10.1103/PhysRevB.47.558}.

\bibitem[{\citenamefont{Bernardi}(2020)}]{Bernardi_one_fourth}
\bibinfo{author}{\bibfnamefont{M.}~\bibnamefont{Bernardi}},
  \bibinfo{journal}{Journal of Physics: Condensed Matter}
  \textbf{\bibinfo{volume}{32}}, \bibinfo{pages}{385501}
  (\bibinfo{year}{2020}),
  \urlprefix\url{https://dx.doi.org/10.1088/1361-648X/ab9409}.

\bibitem[{\citenamefont{Gali et~al.}(2009)\citenamefont{Gali, Janz\'en, De\'ak,
  Kresse, and Kaxiras}}]{Gali-delta-SCF}
\bibinfo{author}{\bibfnamefont{A.}~\bibnamefont{Gali}},
  \bibinfo{author}{\bibfnamefont{E.}~\bibnamefont{Janz\'en}},
  \bibinfo{author}{\bibfnamefont{P.}~\bibnamefont{De\'ak}},
  \bibinfo{author}{\bibfnamefont{G.}~\bibnamefont{Kresse}}, \bibnamefont{and}
  \bibinfo{author}{\bibfnamefont{E.}~\bibnamefont{Kaxiras}},
  \bibinfo{journal}{Phys. Rev. Lett.} \textbf{\bibinfo{volume}{103}},
  \bibinfo{pages}{186404} (\bibinfo{year}{2009}),
  \urlprefix\url{https://link.aps.org/doi/10.1103/PhysRevLett.103.186404}.

\bibitem[{\citenamefont{Ivanov et~al.}(2022)\citenamefont{Ivanov, Simoni, Lee,
  Liu, Jhuria, Redjem, Zhiyenbayev, Papapanos, Qarony, Kant\'e
  et~al.}}]{Ivanov_2}
\bibinfo{author}{\bibfnamefont{V.}~\bibnamefont{Ivanov}},
  \bibinfo{author}{\bibfnamefont{J.}~\bibnamefont{Simoni}},
  \bibinfo{author}{\bibfnamefont{Y.}~\bibnamefont{Lee}},
  \bibinfo{author}{\bibfnamefont{W.}~\bibnamefont{Liu}},
  \bibinfo{author}{\bibfnamefont{K.}~\bibnamefont{Jhuria}},
  \bibinfo{author}{\bibfnamefont{W.}~\bibnamefont{Redjem}},
  \bibinfo{author}{\bibfnamefont{Y.}~\bibnamefont{Zhiyenbayev}},
  \bibinfo{author}{\bibfnamefont{C.}~\bibnamefont{Papapanos}},
  \bibinfo{author}{\bibfnamefont{W.}~\bibnamefont{Qarony}},
  \bibinfo{author}{\bibfnamefont{B.}~\bibnamefont{Kant\'e}},
  \bibnamefont{et~al.}, \bibinfo{journal}{Phys. Rev. B}
  \textbf{\bibinfo{volume}{106}}, \bibinfo{pages}{134107}
  (\bibinfo{year}{2022}),
  \urlprefix\url{https://link.aps.org/doi/10.1103/PhysRevB.106.134107}.

\bibitem[{\citenamefont{Montereali et~al.}(2012)\citenamefont{Montereali,
  Bonfigli, Menchini, and Vincenti}}]{Montereali3}
\bibinfo{author}{\bibfnamefont{R.~M.} \bibnamefont{Montereali}},
  \bibinfo{author}{\bibfnamefont{F.}~\bibnamefont{Bonfigli}},
  \bibinfo{author}{\bibfnamefont{F.}~\bibnamefont{Menchini}}, \bibnamefont{and}
  \bibinfo{author}{\bibfnamefont{M.~A.} \bibnamefont{Vincenti}},
  \bibinfo{journal}{Low Temperature Physics} \textbf{\bibinfo{volume}{38}},
  \bibinfo{pages}{779} (\bibinfo{year}{2012}), ISSN \bibinfo{issn}{1063-777X},
  \eprint{https://pubs.aip.org/aip/ltp/article-pdf/38/8/779/15917553/779\_1\_online.pdf},
  \urlprefix\url{https://doi.org/10.1063/1.4740241}.

\bibitem[{\citenamefont{Thiering and Gali}(2017)}]{Gali-JT1}
\bibinfo{author}{\bibfnamefont{G.~m.~H.} \bibnamefont{Thiering}}
  \bibnamefont{and} \bibinfo{author}{\bibfnamefont{A.}~\bibnamefont{Gali}},
  \bibinfo{journal}{Phys. Rev. B} \textbf{\bibinfo{volume}{96}},
  \bibinfo{pages}{081115} (\bibinfo{year}{2017}),
  \urlprefix\url{https://link.aps.org/doi/10.1103/PhysRevB.96.081115}.

\bibitem[{\citenamefont{Ádám Gali}(2019)}]{Gali-JT2}
\bibinfo{author}{\bibnamefont{Ádám Gali}}, \bibinfo{journal}{Nanophotonics}
  \textbf{\bibinfo{volume}{8}}, \bibinfo{pages}{1907} (\bibinfo{year}{2019}),
  \urlprefix\url{https://doi.org/10.1515/nanoph-2019-0154}.

\bibitem[{\citenamefont{Bosi et~al.}(1970)\citenamefont{Bosi, Bussolati, and
  Spinolo}}]{Bosi_lifetimes}
\bibinfo{author}{\bibfnamefont{L.}~\bibnamefont{Bosi}},
  \bibinfo{author}{\bibfnamefont{C.}~\bibnamefont{Bussolati}},
  \bibnamefont{and} \bibinfo{author}{\bibfnamefont{G.}~\bibnamefont{Spinolo}},
  \bibinfo{journal}{Physics Letters A} \textbf{\bibinfo{volume}{32}},
  \bibinfo{pages}{159} (\bibinfo{year}{1970}), ISSN \bibinfo{issn}{0375-9601},
  \urlprefix\url{https://www.sciencedirect.com/science/article/pii/0375960170902525}.

\bibitem[{\citenamefont{Lany}(2011)}]{Lany}
\bibinfo{author}{\bibfnamefont{S.}~\bibnamefont{Lany}},
  \bibinfo{journal}{physica status solidi (b)} \textbf{\bibinfo{volume}{248}},
  \bibinfo{pages}{1052} (\bibinfo{year}{2011}),
  \urlprefix\url{https://onlinelibrary.wiley.com/doi/abs/10.1002/pssb.201046274}.

\bibitem[{\citenamefont{Lany and Zunger}(2009)}]{Zunger_Koopmans}
\bibinfo{author}{\bibfnamefont{S.}~\bibnamefont{Lany}} \bibnamefont{and}
  \bibinfo{author}{\bibfnamefont{A.}~\bibnamefont{Zunger}},
  \bibinfo{journal}{Phys. Rev. B} \textbf{\bibinfo{volume}{80}},
  \bibinfo{pages}{085202} (\bibinfo{year}{2009}),
  \urlprefix\url{https://link.aps.org/doi/10.1103/PhysRevB.80.085202}.

\bibitem[{\citenamefont{Kurobori et~al.}(1988)\citenamefont{Kurobori, Kanasaki,
  Imai, and Takeuchi}}]{Kurobori_lifetimes}
\bibinfo{author}{\bibfnamefont{T.}~\bibnamefont{Kurobori}},
  \bibinfo{author}{\bibfnamefont{T.}~\bibnamefont{Kanasaki}},
  \bibinfo{author}{\bibfnamefont{Y.}~\bibnamefont{Imai}}, \bibnamefont{and}
  \bibinfo{author}{\bibfnamefont{N.}~\bibnamefont{Takeuchi}},
  \bibinfo{journal}{Journal of Physics C: Solid State Physics}
  \textbf{\bibinfo{volume}{21}}, \bibinfo{pages}{L397} (\bibinfo{year}{1988}),
  \urlprefix\url{https://dx.doi.org/10.1088/0022-3719/21/12/009}.

\end{thebibliography}

\end{document}